\documentclass[citeautoscript,aps,prb,floatfix,showpacs,twocolumn,superscriptaddress,longbibliography]{revtex4-2}
\usepackage{amsmath}
\usepackage{graphicx}
\usepackage{latexsym}
\usepackage{bbold}
\usepackage{accents}
\usepackage{bm}
\usepackage{color}
\usepackage{enumitem} 
\usepackage{ragged2e} % For fine control over justification
\usepackage{subcaption}
\usepackage{float} % Ensures proper float handling
\usepackage{comment}
\usepackage{enumitem}
\usepackage{multirow}

\usepackage{natbib}

\usepackage{modiagram}
\usepackage[version=3]{mhchem}
\usepackage{chemfig}
\usepackage{tikzorbital}

\definecolor{purple2}{rgb}{0.8, 0.0,1}
\definecolor{forestgreen}{rgb}{0.13, 0.55, 0.13}
\definecolor{new_green}{rgb}{0.4, 0.69, 0.2}

\usepackage[%
  colorlinks=true,
  urlcolor=blue,
  linkcolor=blue,
  citecolor=blue
]{hyperref}

\usepackage{tikz}
\usepackage{modiagram}

\begin{document}

\title{Systematic generation of electron models for Second-Principles Density Functional Theory Methods}

\author{ Nayara Carral-Sainz }
\email{nayara.carral@unican.es}
\affiliation{ Departamento de Ciencias de la Tierra y
	F\'{\i}sica de la Materia Condensada, Universidad de Cantabria,
	Avenida de los Castros s/n, 39005 Santander, Spain.}

\author{ Toraya Fern\'andez-Ruiz }
\affiliation{ Departamento de Ciencias de la Tierra y
	F\'{\i}sica de la Materia Condensada, Universidad de Cantabria,
	Avenida de los Castros s/n, 39005 Santander, Spain.}

\author{ Jorge \'I\~niguez-Gonz\'alez }
\affiliation{Materials Research and Technology Department, Luxembourg Institute of Science and Technology, 5 avenue des Hauts-Fourneaux, L-4362 Esch/Alzette, Luxembourg}
\affiliation{Department of Physics and Materials Science, University of Luxembourg, 41 Rue du Brill, Belvaux L-4422, Luxembourg}

\author{ Javier Junquera }
\affiliation{ Departamento de Ciencias de la Tierra y
	F\'{\i}sica de la Materia Condensada, Universidad de Cantabria,
	Avenida de los Castros s/n, 39005 Santander, Spain.}

\author{ Pablo Garc\'{\i}a-Fern\'andez }
\email{garciapa@unican.es}
\affiliation{ Departamento de Ciencias de la Tierra y
	F\'{\i}sica de la Materia Condensada, Universidad de Cantabria,
	Avenida de los Castros s/n, 39005 Santander, Spain.}

\date{\today}

\begin{abstract}
We present a systematic, quasi-automated methodology for generating electronic models in the framework of second-principles density functional theory (SPDFT). This approach enables the construction of accurate and computationally efficient models by deriving all necessary parameters from first-principles calculations on a carefully designed training set. A key feature of our method is the enforcement of space group symmetries, which reduces both the number of independent parameters and the required computational effort. The formalism includes improved treatments of one-electron Hamiltonians, electron-lattice coupling—through both linear and quadratic terms—and electron-electron interactions, enabling accurate modeling of structural and electronic responses. We apply the methodology to SrTiO$_{3}$ and LiF, materials representative of transition-metal perovskites and wide-band-gap insulators, respectively. In both cases, the resulting models reproduce DFT reference data with high fidelity across various atomic configurations and charge states. Our results validate the robustness of the approach and highlight its potential for simulating complex phenomena such as polarons and excitons. This work lays the foundation for extending SPDFT to real-time simulations of optoelectronic properties and further integration with machine-learning methods.
\end{abstract}

\maketitle

\section{Introduction}

First-principles simulations have become an essential tool for studying physical and chemical systems, encompassing atoms, molecules, and solids~\cite{Martin_2004,Kohanoff_2006}. These computational techniques provide a wealth of information at the atomic level that is often complementary to experimental data. Moreover, they are particularly valuable for interpreting phenomena at the microscopic level, offering quantitative insights into the relative importance of competing mechanisms in complex systems.

In condensed matter physics, and to a lesser extent in molecular physics, density functional theory (DFT) has been the method of choice for the past three decades due to its balance between accuracy and computational cost~\cite{Parr_book,Dreizler_book}. 
While quantum chemistry methods~\cite{Szabo_book} can yield highly accurate energy calculations, they are typically restricted to small molecular systems or limited-size simulation cells. In contrast, DFT can handle hundreds or even thousands of atoms under periodic boundary conditions, making it highly suitable for the study of materials~\cite{Bowler_2010}.

However, many practical problems require climbing the multi-scale ladder in materials simulations, as different physical phenomena emerge at distinct length and time scales.
Addressing these challenges demands a seamless connection between atomistic and continuum descriptions.
Despite ongoing advancements aimed at improving its efficiency and expanding its applicability, DFT remains computationally prohibitive for reaching the mesoscale, where collective behaviors and emergent properties become significant. 
Moreover, DFT simulations are often restricted to zero or very low temperatures due to the high computational cost of capturing finite-temperature dynamics with an explicit quantum mechanical treatment. 
In these multi-scale approaches, it is often desirable to retain an explicit electronic description because the relevant physics might be dictated by their behaviour. It is also important to maintain a high accuracy, as many key properties—such as phase transitions, symmetry breaking, and collective interactions—depend on the fine balance of quantum and statistical effects across different scales~\cite{anderson_science72,Hwang-12,Nagaosa_pscr12}. The description of systems exhibiting such novel properties—such as exotic conducting states~\cite{ohtomo_nat04}, or non-trivial topological phases of magnetization~\cite{nagaosa_natnano13} or polarization~\cite{das_nature19,Junquera-23}—often requires the use of very large simulation cells, exceeding the computational capabilities of standard DFT methods. Furthermore, some phenomena demand extensive statistical sampling, as in Monte Carlo simulations~\cite{Binder_book}, or long-time molecular dynamic simulations involving large systems. Notable examples include the dynamics of polarons or excitons, which can hop or tunnel between lattice sites while interacting with other degrees of freedom such as lattice vibrations, impurities, or magnetic moments~\cite{schnakenberg_pssb68,asadi_natcomm13,castner_jpcs57,kanzig_jpcs59}.
Certainly, these problems are of significant applied interest, as these phenomena play an important role in the efficiency of battery materials~\cite{garcialastra_jpcc13,ong_prb12}, photovoltaic cells~\cite{guo_advmat20} or light-emitting diodes~\cite{guo_advmat22} to name a few examples. 
Thus, accurately predicting the outcome of such experiments requires computational approaches that extend beyond standard DFT.

In a previous work~\cite{pgf_prb16}, we introduced {\sl second-principles DFT} (SPDFT) calculations, designed to address large-scale problems including both atomic and electronic degrees of freedom on the same footing in systems with many atoms and extended simulation times. This method constructs general models capable of describing diverse properties, mechanical, structural, electrical, or magnetic, while maintaining a computational cost much lower than that of standard DFT calculations. SPDFT models can capture essential physical properties such as total energy, electronic bands, and magnetic moments for systems containing tens or even hundreds of thousands of atoms. Importantly, this approach retains predictive accuracy, as its parameters are systematically derived from DFT calculations and can, in principle, be refined to converge toward first-principles results.

Since our initial publication~\cite{pgf_prb16}, machine-learning (ML) models in condensed matter physics and chemistry have proliferated~\cite{schmidt_npj19,baum_jcim21}. 
Universal deep neural network approaches to represent the DFT Hamiltonian of crystalline materials, aiming to bypass the computationally demanding self-consistent field iterations of DFT and substantially improve the efficiency of \textit{ab initio} electronic-structure calculations, are already available~\cite{Li-22,website_deepH}.
Machine learning and artificial intelligence models have proven to be highly accurate in condensed matter physics, capable of predicting material properties, phase transitions, and electronic structures in large systems with remarkable precision. However, the complexity of these models, often involving deep neural networks or high-dimensional feature spaces, makes them inherently opaque. Unlike traditional physics-based approaches, where analytical equations offer clear insights into underlying mechanisms, ML-driven models function as ``black boxes'', providing results without a transparent explanation of the reasoning behind them. This lack of interpretability poses a challenge, as scientists may obtain precise predictions but struggle to extract fundamental physical principles or causal relationships from the model’s inner workings.
Moreover, ML-based approaches are often tailored to specific properties (such as energy gap predictions~\cite{zhuo_jpcl18}) or materials (for instance, the mechanical properties of zeolites~\cite{evans_chemmat17}). 
By contrast, SPDFT offers a physics-based framework that applies to a broader range of systems and properties, making it a valuable complement to data-driven methodologies.

In Ref.~\cite{pgf_prb16}, we also proposed an algorithm to generate the electronic component of SPDFT models by systematically filtering one-electron Hamiltonians expressed in a Wannier Function (WF) basis~\cite{Marzari-12}. These Hamiltonians were obtained from DFT calculations on a loosely defined training set, allowing users to incorporate specific configurations relevant to the system under study. However, this approach had several limitations:
(i) It did not explicitly enforce the symmetry of the system; neither the way the parameters were obtained nor the training set force the parameters to be consistent with the space group of the reference geometry, leading to discrepancies in equivalent parameters and potential errors in band structure representation.
(ii) The lack of a formally defined training set made it difficult to compare different models, as they contained non-equivalent terms derived from disparate DFT calculations.
(iii) Model generation required an extensive number of DFT calculations (potentially tens of thousands), incurring significant computational costs, which contradicted the efficiency goals of the method.
(iv) There was no systematic way to assess the quality of a given model, making it challenging to refine and improve parameter sets.

In this work, we address these issues by developing an improved, largely-automated SPDFT model generation algorithm. The method requires users to specify only a few computational thresholds, which are then systematically employed to construct the training set and evaluate the quality of the resulting model. As we will demonstrate, this approach yields models that perform well on standardized test sets.

While the present work is primarily devoted to improving the methodology for constructing second-principles models, it does not include a comprehensive validation in more complex or application-driven scenarios involving the prediction of non-trivial physical properties. Demonstrating such capabilities remains a central objective of the SPDFT framework and was already exemplified in our earlier work~\cite{pgf_prb16}. Nonetheless, we have verified that the models generated using the methodology presented here are capable of capturing subtle and intricate physical effects. These results will be discussed in future publications, highlighting the predictive potential of SPDFT across a broad range of phenomena.
For instance, in Sec.~\ref{sec:lif}, we evaluate the performance of the model for LiF using a standardized test set of DFT calculations designed to quantify model accuracy under controlled distortions. Beyond this, we have confirmed—though it lies outside the scope of the current manuscript—that this model, when combined with hybrid functionals and the recently developed time-dependent extension of SPDFT, can successfully describe optical properties such as excitons, which are governed by strong electron-hole interactions. These applications further underscore the utility of the approach in addressing realistic and technologically relevant material responses.

The paper is organized as follows; 
In the following section, Sec.~\ref{sec:overview}, we provide an overall view of SPDFT~\cite{pgf_prb16},
and use it to introduce the notation and the parameters that need to be determined.
We then introduce several changes in the SPDFT Hamiltonian in Sec.~\ref{spdft_hamiltonian} 
that allow for significant improvements in the description of 
metals and electron-lattice interactions.
The main goal of this work, the method used to obtain the parameters 
of the effective Hamiltonian from first-principles, is described 
in Sec.~\ref{sec_electron}.
This includes the procedure to obtain the Wannier functions used
to create the models (Sec.~\ref{sec:mani}), the symmetrization of the parameters (Sec.~\ref{sec:symm})
and the way in which the parameters themselves are obtained and checked (Secs.~\ref{sec:param} and \ref{sec:verify}).
We then move to provide some computational details, Sec.~\ref{sec:computational},
associated to the various applications developed in Sec.~\ref{sec:results}.
These include the discussion of the models of SrTiO$_3$, Sec.~\ref{sec:STO}, 
an important perovskite material with significant electron-lattice coupling; and LiF,
Sec.~\ref{sec:lif},
a rock-salt material whose optical properties display strongly 
bound excitons, underscoring the importance of electron-hole coupling
in the system. 
We finally present our conclusions and discuss the future applications
of second-principles calculations in Sec.~\ref{sec:conclusions}.

\section{Overview of the method}\label{sec:overview}

The SPDFT method~\cite{pgf_prb16} is founded on Taylor expansions of both lattice and electronic energies about reference configurations.  
In this context, two fundamental concepts are introduced.

The first is the \emph{reference (not relaxed) atomic geometry} (hereafter RAG), which denotes a particular arrangement of atomic nuclei taken as the baseline for describing other configurations.  
Assuming the Born–Oppenheimer approximation, the energy of the lattice is expressed as a Taylor expansion of the potential energy surface (PES) around the RAG, formulated as a function of all atomic degrees of freedom and the strain tensor.  
The RAG is typically chosen as a high-symmetry equilibrium point on the system’s PES.  
Any arbitrary crystal configuration may then be represented by expressing the atomic positions, $\vec{r}_{\bm{\lambda}}$, as a distortion of the RAG through the following relation,

\begin{align}
\vec{r}_{\bm{\lambda}}= (\mathbb{1}+\overleftrightarrow{\eta})\left(\vec{R}_{\Lambda}+
    \vec{\tau}_{\lambda} \right) +
    \vec{u}_{\bm{\lambda}},
    \label{eq:atom_pos}
\end{align}

\noindent where the bold Greek subscript $\bm{\lambda}$ comprises both the atomic index within the primitive cell ($\lambda$) and the corresponding Bravais lattice vector $\vec{R}_\Lambda$.  
Thus, in the RAG, the atom resides at position $\vec{R}_{\Lambda} + \vec{\tau}_\lambda$, and its total displacement from the reference geometry is composed of a homogeneous strain, $\overleftrightarrow{\eta}$, which affects the entire lattice, and a local displacement within the unit cell, $\vec{u}_{\bm{\lambda}}$.  

\begin{figure}[t] 
    \centering    
    \includegraphics[width=\linewidth]{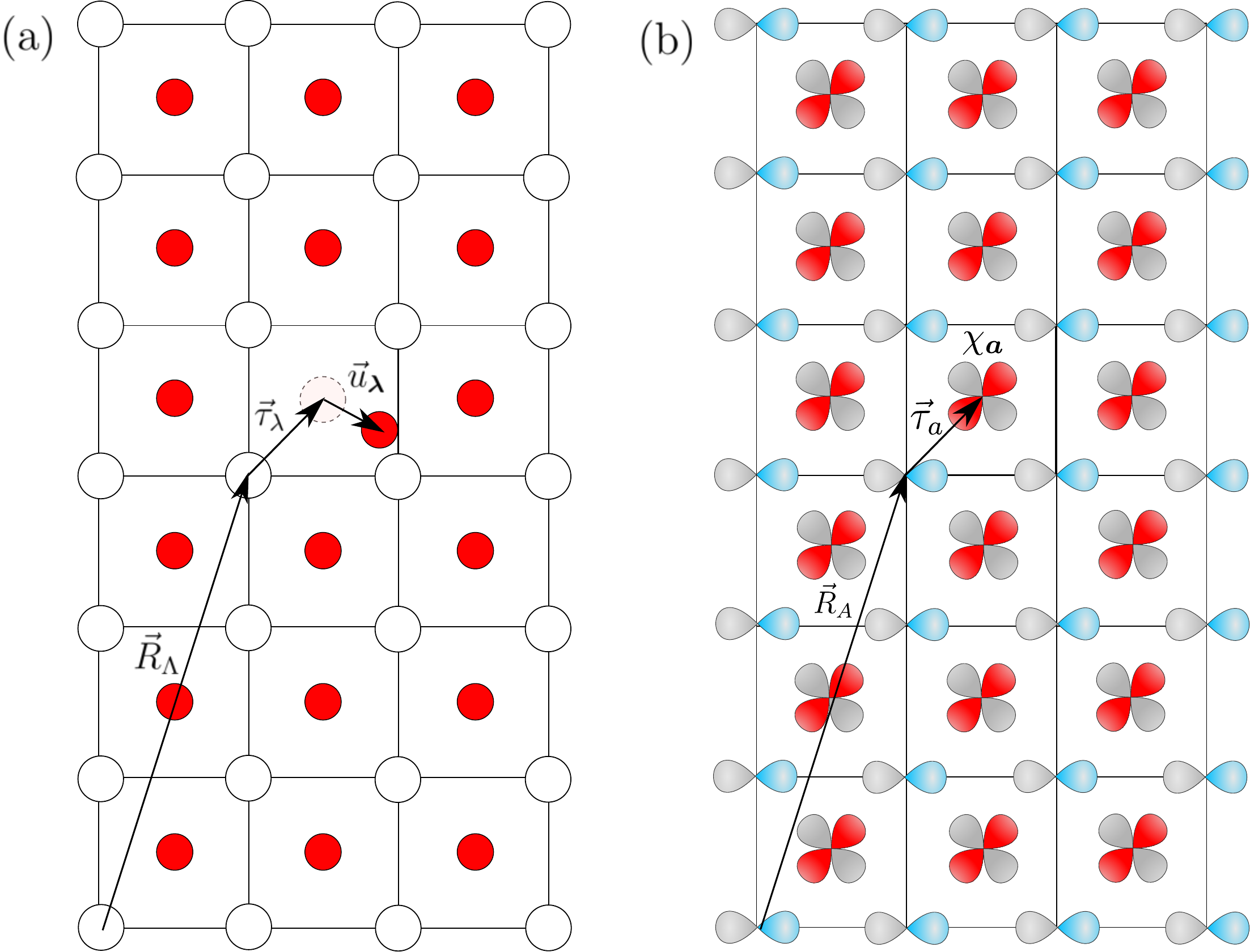}
    \caption{\justifying
    Atomic and Wannier function positions.
    (a) Schematic representation of the vectors relevant to the description of atomic positions within the RAG structure, including atomic displacements.
    (b) Representation of the vectors associated with the spatial locations of the Wannier functions.
     Figure adapted from \cite{wojdel_jpcm13}. 
    }
    \label{fig:wide}
\end{figure}

The second foundational concept is the \emph{reference electron density} (hereafter RED), denoted $n_{0}(\vec{r})$, which is defined \emph{for each atomic configuration}.  
For non-magnetic insulating systems, the RED may correspond to the ground-state self-consistent charge density. Conversely, in metallic and magnetic systems, it can be associated with a convenient though unphysical (i.e., non-observable) reference state~\cite{pgf_prb16}.  
Section~\ref{sec:RED} will present a more detailed discussion on appropriate strategies for selecting the RED in SPDFT.

Many physically relevant scenarios necessitate the modeling of electronic states deviating from the ground state—such as the formation of polarons via charge addition or removal, or the generation of excitons through electron-hole pairs.  
In these cases, the self-consistent charge density must account for both the added, removed and/or excited charge and the induced redistribution of the background electron density due to screening effects.  
The total charge density can thus be decomposed as

\begin{equation}
   n(\vec{r})=n_0(\vec{r})+\delta n(\vec{r}),
   \label{eq:dens_div}
\end{equation}

\noindent where $\delta n(\vec{r})$ denotes the deformation or difference density, which is assumed to be sufficiently small to permit a perturbative treatment.

The electron density is represented using a set of localized, geometry-dependent Wannier functions (WFs), $\chi_{\bm{a}}$.  
Here, the bold Latin subscript $\bm{a}$ encapsulates both the WF index within the primitive cell ($a$) and the Bravais lattice vector of the corresponding unit cell, $\vec{R}_{A}$.  
Using this basis, the total, reference, and deformation densities are expressed as

\begin{equation}
n(\vec{r})=\sum_{\bm{a}\bm{b}} d_{\bm{a}\bm{b}} \chi_{\bm{a}}(\vec{r}) \chi_{\bm{b}}(\vec{r}),
\label{eq:densmat}
\end{equation}
\begin{equation}
n_0(\vec{r})=\sum_{\bm{a}\bm{b}} d_{\bm{a}\bm{b}}^{(0)} \chi_{\bm{a}}(\vec{r}) \chi_{\bm{b}}(\vec{r}),
\label{eq:densmat_ref}
\end{equation}
\begin{equation}
\delta n(\vec{r})=n(\vec{r})-n_0(\vec{r})= \sum_{\bm{a}\bm{b}} D_{\bm{a}\bm{b}} \chi_{\bm{a}}(\vec{r}) \chi_{\bm{b}}(\vec{r}),
\label{eq:densmat_diff}
\end{equation}

\noindent where $d_{\bm{a}\bm{b}}$, $d^{(0)}_{\bm{a}\bm{b}}$, and $D_{\bm{a}\bm{b}}$ denote the corresponding density matrices in the geometry-dependent WF basis.  
The difference density matrix is defined as

\begin{equation}
D_{\bm{a}\bm{b}} = d_{\bm{a}\bm{b}}-d_{\bm{a}\bm{b}}^{(0)},
\label{eq:diff_den}
\end{equation}

\noindent and is assumed to be, locally, much smaller than the total number of electrons in that region of the system

\begin{equation}
\sum_{\bm{a}\in \Omega} \vert D_{\bm{a}\bm{a}}\vert \ll \sum_{\bm a\in \Omega} d_{\bm{a}\bm{a}}^{(0)}.
\end{equation}

\noindent where $\Omega$ is a region of the system usually containing a few cells of the system. Under this perturbative assumption, the DFT energy can be expanded to low order in $\delta n$, 

\begin{equation}
  E_{\text{DFT}}\approx E^{(0)}+E^{(1)}+E^{(2)}+ \ldots
\end{equation}

The zeroth-order term, $E^{(0)}$, exactly corresponds to the DFT energy evaluated for the RED, $n_{0}(\vec{r})$, and it constitutes the dominant contribution to the total energy.

The first-order term, $E^{(1)}$, describes one-electron excitations and represents the \textit{difference} between the sum of one-electron energies in the perturbed system and the sum of one-electron energies in the reference configuration.  
As a difference, $E^{(1)}$ can be written as a function of the deformation density matrix and is generally a small quantity, enabling an accurate computational treatment.

The second-order term, $E^{(2)}$, accounts for the screening of two-electron interactions via exchange and correlation effects.

The final expression for the total energy, derived in Ref.~\cite{pgf_prb16}, can be written in the Wannier basis as

\begin{align}
   E = & E^{(0)} + 
         \sum_{\bm{a}\bm{b}} 
           \left(
                 D_{\bm{a}\bm{b}}^\uparrow+D_{\bm{a}\bm{b}}^\downarrow
           \right)
           \gamma_{\bm{a}\bm{b}}%^\text{sh} 
   \nonumber \\
   &+ \frac{1}{2}
     \sum_{\bm{a}\bm{b}}
     \sum_{\bm{a}^\prime\bm{b}^\prime} 
   \left\{
        \left(D_{\bm{a}\bm{b}}^\uparrow + D_{\bm{a}\bm{b}}^\downarrow\right)
        \left(D_{\bm{a}^\prime\bm{b}^\prime}^\uparrow + D_{\bm{a}^\prime\bm{b}^\prime}^\downarrow\right)
   \right.
      U_{\bm{a}\bm{b}\bm{a}^\prime\bm{b}^\prime}
   \nonumber \\
   &\left.
      -\left(D_{\bm{a}\bm{b}}^\uparrow - D_{\bm{a}\bm{b}}^\downarrow \right)
       \left(D_{\bm{a}^\prime\bm{b}^\prime}^\uparrow - D_{\bm{a}^\prime\bm{b}^\prime}^\downarrow\right)
      I_{\bm{a}\bm{b}\bm{a}^\prime\bm{b}^\prime}
   \right\},
\label{eq:totalenergy1updn}
\end{align}

\noindent where we use arrows to denote the spin-up and spin-down parts of the difference density.
This expression highlights that the difference between the the total energy and that of the reference system, $E^{(0)}$, depends solely on the difference density matrix, $D_{\bm{a}\bm{b}}$, and a set of integrals. These include the effective one-electron tight-binding-like Hamiltonian matrix elements, $\gamma_{\bm{a}\bm{b}}$, as well as their responses to changes in charge density and spin polarization, denoted by $U_{\bm{a}\bm{b}\bm{a}^\prime\bm{b}^\prime}$ and $I_{\bm{a}\bm{b}\bm{a}^\prime\bm{b}^\prime}$, respectively.  
Formally, this expression closely resembles the energy expression in Hartree-Fock theory; however, the integrals, derived within the framework of density functional theory (DFT), inherently incorporate electron correlation effects.
Moreover, due to the spatial localization of Wannier functions, the integrals are short-ranged  (save electrostatic couplings \cite{pgf_prb16}), thereby enabling efficient large-scale simulations.

The band structure and Bloch wavefunctions can be obtained from the one-particle Hamiltonian, whose expression for the spin channel \( s \) is given by

\begin{align}
   h_{\bm{a}\bm{b}}^s & = \left\langle\chi_{\bm{a}}^{s} \right\vert \hat{h} \left\vert \chi_{\bm{b}}^{s}\right\rangle\nonumber\\
                      & = \gamma_{\bm{a}\bm{b}} + 
      \sum_{\bm{a^\prime b}^\prime} &
           \left[
               \left(
                  D_{\bm{a}^\prime \bm{b}^\prime}^s+
                  D_{\bm{a}^\prime \bm{b}^\prime}^{-s}
               \right)
               U_{\bm{a b a^\prime b}^\prime}-
           \right.
           \nonumber \\
      &&
           \left.
               \left(
                  D_{\bm{a}^\prime\bm{b}^\prime}^{s}-
                  D_{\bm{a}^\prime\bm{b}^\prime}^{-s}
               \right)
               I_{\bm{ab a^\prime b}^\prime}
           \right].          
    \label{eq:honeR}
\end{align}

This concludes the summary of the SPDFT formalism.  
We emphasize that, although certain components are reminiscent of other methods—such as tight-binding DFT~\cite{porezag_prb95,matthew_prb89,elstner_prb98}—the key distinction lies in the specific definitions of the RED and the zeroth-order energy, $E^{(0)}$.  
In SPDFT, $E^{(0)}$ is constructed via a polynomial expansion of the PES to arbitrary order, as detailed in Ref.~\cite{wojdel_jpcm13} and Ref.~\cite{gonze2020}.
Importantly, because the method takes the RED and its energy as a given, the SPDFT electronic model does not need to account for such as quantity (which essentially amounts to the cohesive energy of the system), which allows for relatively simple and computationally light models.

Extending on this last point, because both lattice and electronic energies are expanded about the RAG and RED, respectively, the formalism inherently assumes that no significant bond-topology transformations take place.  
For instance, the method is suitable for describing structural phase transitions such as the ferroelectric transitions in BaTiO$_3$ or PbTiO$_3$, where the underlying perovskite topology is preserved. However, it is not applicable to processes involving bond breaking, such as ionic diffusion or melting.

\section{Improvements on the SPDFT Hamiltonian}
\label{spdft_hamiltonian}

Since the original publication of the SPDFT method~\cite{pgf_prb16}, we have identified several ways to overcome some of its limitations. In particular, we have implemented a number of important modifications to the form of the Hamiltonian, which are detailed below.

\subsection{Reference density}
\label{sec:RED}

In the original work~\cite{pgf_prb16}, the reference electron density (RED) was defined using a geometry-independent, constant diagonal matrix:

\begin{align}\label{eq:old_refdenmat}
  d_{\bm{a}\bm{b}}^{(0)}=o_{\bm a} \delta_{\bm{a}\bm{b}},
\end{align}

\noindent where $o_{\bm a}$ denotes the number electrons on the corresponding WF. This density matrix is periodic in the sense that $d_{\bm{a}\bm{b}}^{(0)} = d_{\bm{a}^\prime\bm{b}^\prime}^{(0)}$ if 
$(\bm{a}^\prime,\bm{b}^\prime) \equiv (\bm{a}^\prime,\bm{b}^\prime)$ with ``$\equiv$'' meaning equivalent by translation. 
This definition, while practical from a computational point of view, has significant shortcomings. 
The total density matrix of any system (insulating or metal) needs to be idempotent. 
In the case of non-magnetic insulators where the occupation of each WF can be chosen to be 0 or 1 and both spin-channels are completely equivalent, Eq.~(\ref{eq:old_refdenmat}) allows an exact description of the ground state and has the advantage of not changing with the geometry unless the system metallizes, i.e. the same states are always occupied (or empty).
However, this approach becomes insufficient for metals and magnetic insulators, where the occupation of the WFs in the RED,  Eq.~(\ref{eq:old_refdenmat}), may be fractional and off-diagonal elements of the density matrix ($d_{\bm{a}\bm{b}}$ with $\bm{a}\neq\bm{b}$) need to be non-zero in the ground state to fulfill idempotency. Accurately capturing the ground state density matrix in such systems therefore requires a self-consistent approach, which can be both time-consuming and technically demanding.

In the new implementation of SPDFT, the reference density matrix is no longer constrained to be diagonal, although we force it to remain constant (in the sense that the subsets of zero and non-zero  components remain mutually exclusive throughout) with respect to the geometry. This generalization allows us to select the RED as the ground state density matrix of the reference atomic geometry (RAG) in both insulating and metallic systems. 

For insulators, if the RED is diagonal, this choice also describes the electronic ground state exactly across different geometries. In contrast, when the RED is not diagonal, typically in metals where band occupation depends on the geometry, the RED only coincides with the ground state at the RAG. This means that, while in non-magnetic insulators $E^{(0)}$ represents exactly the ground state energy, in metals $E^{(1)}$ and $E^{(2)}$ need to be included outside the RAG. This behavior ultimately reflects the scattering of the electrons by the lattice in metals.  Nonetheless, the current approach leads to smaller variations in the density matrix and enables a more accurate description of its geometry dependence, which is captured through electron-lattice interaction elements (see Sec.~\ref{sec:ellat}).

\subsection{Linear electron-lattice Hamiltonian}\label{sec:ellat}

An important factor in many physical phenomena is the dependence of the electronic
properties with the geometry of the system. 
In SPDFT this effect is taken into account considering that the one-electron Hamiltonian elements, $\gamma_{\bm{a}\bm{b}}$, explicitly depend on the atomic configuration. 
More precisely, these Hamiltonian matrix elements at the RAG are expanded considering linear and quadratic corrections with respect to the change of the atomic geometry around the RAG. 
In this way, the one-electron Hamiltonian matrix elements were written in the original version of the method as

 \begin{align}
    \gamma_{\bm{a}\bm{b}} = \gamma^{\text{RAG}}_{\bm{a}\bm{b}} & +        
       \sum_{\bm{\lambda}\bm{\upsilon}} 
       \left[
       -\vec{f}_{\bm{a}\bm{b},\bm{\lambda}\bm{\upsilon}}^T
        \delta \vec{r}_{\bm{\lambda}\bm{\upsilon}} + 
       \right.
       \nonumber \\
       &+\left.
       %\sum_{\bm{\lambda}^\prime\bm{\upsilon}^\prime} 
       \delta \vec{r}_{\bm{\lambda}\bm{\upsilon}}^T       
       \overleftrightarrow{g}_{\bm{a}\bm{b},\bm{\lambda}\bm{\upsilon}}
       \delta \vec{r}_{\bm{\lambda}\bm{\upsilon}}+...
       \right], 
    \label{eq:gamma_ellat_old}       
 \end{align}
\noindent where $\vec{f}_{\bm{a}\bm{b},\bm{\lambda}\bm{\upsilon}}$ and 
$\overleftrightarrow{g}_{\bm{a}\bm{b},\bm{\lambda}\bm{\upsilon}}$
are, respectively, the first and second-order electron-lattice coupling parameters. 
$\delta \vec{r}_{\bm{\lambda}\bm{\upsilon}}$ quantifies the relative displacements of atoms 
$\bm{\lambda}$ and $\bm{\upsilon}$,

 \begin{equation}
    \delta \vec{r}_{\bm{\lambda}\bm{\upsilon}} =
    \overleftrightarrow{\eta}\left(\vec{R}_{\Upsilon}-\vec{R}_{\Lambda}+
    \vec{\tau}_{\upsilon}-\vec{\tau}_{\lambda} \right) +
    \vec{u}_{\bm{\upsilon}}-\vec{u}_{\bm{\lambda}}.
 \end{equation}

\noindent This approximation has the advantage that it automatically complies with the acoustic sum rule (ASR), due to the fact that $\delta \vec{r}_{\bm{\lambda}\bm{\upsilon}}$ does not change when all the atoms in the solid are displaced by the same amount. 

During our attempts to systematically fit the electron-lattice coupling parameters \cite{pgf_prb16}, we identified significant inconsistencies in the values of $\vec{f}_{\bm{a}\bm{b},\bm{\lambda}\bm{\upsilon}}$ when using two different sets of reference calculations. This issue becomes apparent when comparing Eq.~(\ref{eq:gamma_ellat_old}) \cite{pgf_prb16} with a direct Taylor expansion of the one-electron Hamiltonian matrix elements with respect to the atomic displacements

 \begin{align}
    \gamma_{\bm{a}\bm{b}} = \gamma^{\text{RAG}}_{\bm{a}\bm{b}} & -        
       \sum_{\bm{\lambda}} 
       \vec{f}_{\bm{a}\bm{b},\bm{\lambda}}^T
        \delta\vec{r}_{\bm{\lambda}} + 
       \sum_{\bm{\lambda}\bm{\upsilon}} 
       \delta \vec{r}_{\bm{\lambda}}^T       
       \overleftrightarrow{g}_{\bm{a}\bm{b},\bm{\lambda}\bm{\upsilon}}
       \delta \vec{r}_{\bm{\upsilon}}+... ,
    \label{eq:gamma_ellat_new}       
 \end{align}

\noindent where the atomic displacements $\delta\vec{r}_{\bm{\lambda}}$ are given by

\begin{equation}
    \delta\vec{r}_{\bm{\lambda}} = \overleftrightarrow{\eta}\left(\vec{R}_{\Lambda} + \vec{\tau}_{\bm{\lambda}}\right) + \vec{u}_{\bm{\lambda}}.
\end{equation}

\noindent It is clear that $\delta\vec{r}_{\bm{\lambda\upsilon}}$ and $\delta\vec{r}_{\bm{\lambda}}$ are related through
\begin{equation} \label{eq:delta_relations}
    \delta \vec{r}_{\bm{\lambda}\bm{\upsilon}} = \delta\vec{r}_{\bm{\upsilon}} - \delta\vec{r}_{\bm{\lambda}}.
\end{equation}

\noindent Moreover, the contribution of the linear part of Eqs.~(\ref{eq:gamma_ellat_old}) and (\ref{eq:gamma_ellat_new}) needs to be the same. Writing this equivalence, taking into account Eq.~(\ref{eq:delta_relations}), and noticing that $\vec{f}_{\bm{ab},\bm{\lambda\upsilon}}=-\vec{f}_{\bm{ab},\bm{\upsilon\lambda}}$ we find,
\begin{align}
\vec{f}_{\bm{a}\bm{b},\bm{\lambda}}= -\sum_{\bm{\upsilon}} 2\vec{f}_{\bm{a}\bm{b},\bm{\lambda}\bm{\upsilon}}.
\end{align}

%\noindent Examining the linear terms in Eq.~(\ref{eq:gamma_ellat_new}), it is evident that there are $3N_{\mathrm{at}}$ independent parameters to be fitted for a system containing $N_{\mathrm{at}}$ atoms. In contrast, the earlier expression [Eq.(\ref{eq:gamma_ellat_old})] requires fitting $(3N_{\mathrm{at}})^2 - 3N_{\mathrm{at}}$ parameters, which is a substantially larger number. The reduction by $3N_{\mathrm{at}}$ arises because the diagonal terms in the pairwise displacement vanish, i.e., $\delta \vec{r}_{\bm{\lambda}\bm{\lambda}} = 0$.
%
%As demonstrated in Appendix~\ref{sec:linear_constants}, the one-body linear coefficients can be obtained from the two-body ones via the relation {\color{red} Include an appendix to explain this.}{\color{green} Done.}
%

\noindent Similarly to what happens in lattice models \cite{wojdel_jpcm13}, this indicates that the two-body coefficients $\vec{f}_{\bm{a}\bm{b},\bm{\lambda}\bm{\upsilon}}$ are underdetermined: there are far more of these parameters than corresponding one-body coefficients $\vec{f}_{\bm{a}\bm{b},\bm{\lambda}}$. Consequently, the two-body representation is not uniquely defined and lacks robustness across different reference data sets.
Therefore, Eq.~(\ref{eq:gamma_ellat_new}) provides a more consistent and physically transparent starting point for constructing second-principles models than Eq.~(\ref{eq:gamma_ellat_old}).
We would finally like to stress that this choice is adequate only for linear  terms. In the case of quadratic terms the representation in terms of coordinates or differences of coordinates is rather equivalent while the latter should be favored if the inclusion of anharmonic terms is necessary. 

A key challenge when employing Eq.~(\ref{eq:gamma_ellat_new}) is ensuring that the acoustic sum rule is properly satisfied. By enforcing the physical requirement that the total energy remains invariant under a rigid translation of the entire lattice, one obtains the constraint

\begin{align}
\sum_{\bm{\lambda}} \vec{f}_{\bm{a}\bm{b},\bm{\lambda}} = 0. \label{eq:ASR_linear}
\end{align}

\begin{figure}[H]
    \centering
    \includegraphics[width=\linewidth]{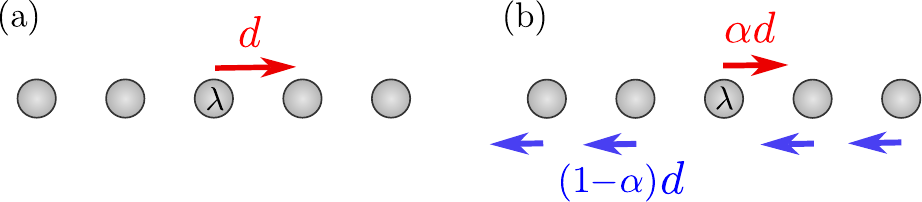}
    \caption{(a) Atoms of an equispaced one-dimensional linear chain, where the atom \(\bm{\lambda}\) has been displaced by a distance \(d\) from its equilibrium position.  
(b) The geometry in [a] is equivalent to a linear chain in which the atom \(\bm{\lambda}\) has been displaced by a distance \(\alpha d\), while the remaining atoms have been moved by a distance \((1-\alpha)d\) in the opposite direction.}
    \label{fig:disp_alpha}
\end{figure}

\noindent The individual coefficients $\vec{f}_{\bm{a}\bm{b},\bm{\lambda}}$ can, in principle, be computed from first-principles by interpreting them as the negative first derivative of $\gamma_{\bm{a}\bm{b}}$ with respect to atomic displacements and evaluating them using finite differences. A straightforward application of this approach—displacing a single atom $\bm{\lambda}$ by a small distance $d$ along a chosen direction, as depicted in Fig.~\ref{fig:disp_alpha}(a),  and extracting the derivative—results in noticeable violations of Eq.~(\ref{eq:ASR_linear}) due to numerical errors that reflect a certain neglect of global translational invariance.

A more robust finite-difference strategy involves displacing atom $\bm{\lambda}$ by a distance $\alpha d$ in a given direction while simultaneously shifting all other atoms in the simulation cell by a distance $(1-\alpha)d$ in the opposite direction, as represented in Fig.~\ref{fig:disp_alpha}(b). Here, $\alpha$ is a real parameter between 0 and 1. Since the relative displacements among atoms are preserved, the computed value of $\vec{f}_{\bm{a}\bm{b},\bm{\lambda}}$ remains unchanged. 
This observation enables a scheme in which the ASR is strictly enforced while still extracting the electron-lattice linear coefficients from finite differences, denoted as $\vec{f}_{\bm{a}\bm{b},\bm{\lambda}}^{\hspace{0.5mm}\mathrm{fd}}$. To implement this, (i) we first define $\alpha = (N_{\mathrm{at}} - 1)/N_{\mathrm{at}}$, where $N_{\mathrm{at}}$ is the total number of atoms in the simulation cell. (ii) For each atom $\bm{\lambda}$, we then assign a linear electron-lattice coupling constant $\vec{f}_{\bm{a}\bm{b},\bm{\lambda}} = \alpha \vec{f}_{\bm{a}\bm{b},\bm{\lambda}}^{\hspace{0.5mm}\mathrm{fd}}$. (iii) The remaining atoms $\bm{\upsilon} \neq \bm{\lambda}$ are each assigned a value $\vec{f}_{\bm{a}\bm{b},\bm{\upsilon}} = -(1 - \alpha) \vec{f}_{\bm{a}\bm{b},\bm{\lambda}}^{\hspace{0.5mm}\mathrm{fd}}$.
It is straightforward to verify that these assignments satisfy Eq.~(\ref{eq:ASR_linear}). Moreover, the resulting coefficients correctly reproduce the desired linear response associated with the relative displacement of atom $\bm{\lambda}$, as determined from finite differences.

\subsection{Quadratic electron-lattice Hamiltonian}

We now turn our attention to the second-order electron-lattice coupling terms. Notably, the number of independent parameters $\overleftrightarrow{g}_{\bm{a}\bm{b},\bm{\lambda}\bm{\upsilon}}$ appearing in Eq.~(\ref{eq:gamma_ellat_old}) and Eq.~(\ref{eq:gamma_ellat_new}) is identical. In fact, one can readily verify that the values of $\overleftrightarrow{g}_{\bm{a}\bm{b},\bm{\lambda}\bm{\upsilon}}$ in Eq.~(\ref{eq:gamma_ellat_old}) are equal in magnitude but opposite in sign to those in Eq.~~(\ref{eq:gamma_ellat_new}).
Moreover, by examining Eq.~(\ref{eq:gamma_ellat_new}) and enforcing the acoustic sum rule—which requires the total energy to remain invariant under a rigid translation of all atoms—we obtain the following condition on the second-order coupling tensors

\begin{align}
\sum_{\bm{\lambda}\neq\bm{\upsilon}}\overleftrightarrow{g}_{\bm{a}\bm{b},\bm{\lambda}\bm{\upsilon}}=
-\overleftrightarrow{g}_{\bm{a}\bm{b},\bm{\upsilon}\bm{\upsilon}}\label{eq:ref_ASL_quadratic}
\end{align}

\noindent This condition ensures that the second-order expansion respects translational invariance. Importantly, the off-diagonal terms $\overleftrightarrow{g}_{\bm{a}\bm{b},\bm{\lambda}\bm{\upsilon}}$ with $\bm{\lambda} \neq \bm{\upsilon}$ can be directly computed from finite-difference calculations. Consequently, imposing the ASR through Eq.~(\ref{eq:ref_ASL_quadratic}) within second-principles frameworks becomes straightforward and computationally efficient.

\subsection{Electrostatics}
\label{sec:electrostatics}

In Ref.~\cite{pgf_prb16}, the electrostatic contribution was described using local dipoles that depended solely on the atomic geometry. However, in subsequent applications—particularly in the context of optical property calculations~\cite{tfr_prb25}—we found that this treatment is insufficient. Specifically, the purely electronic contribution to the  atomic dipoles must be allowed to evolve dynamically, as external electric fields can induce hybridization between Wannier functions, leading to time-dependent polarization effects. This observation has led us to revise the electrostatic model accordingly.
More details will be given in a forthcoming simulation.

\section{Deriving the Parameters from First-Principles}\label{sec_electron}

In the original SPDFT formulation~\cite{pgf_prb16}, a procedure was outlined for fitting parameters for second-principles models. That approach emphasized the importance of filtering the large volume of data produced by first-principles calculations. However, less attention was given to the construction of the {\it training set}, that is, to defining the specific first-principles simulations needed to parameterize a second-principles model.

As we refined the fitting process, we identified two main issues that limited model accuracy. First, one-electron Hamiltonians expressed in the Wannier function (WF) basis often exhibited small numerical inconsistencies. These resulted in symmetry-related terms being treated as distinct, which in turn led to unphysical features in the second-principles models—such as small band splittings where degeneracies were expected.

Second, the standard procedure for WF construction introduced further limitations. Typically, WFs are obtained in a single run by selecting all relevant bands simultaneously, optimizing for maximal localization. When applied, for example, to the valence and conduction bands of an insulator, the resulting WFs mix Bloch states from both manifolds. This mixing leads to a ground-state density matrix (see Sec.~\ref{sec:RED}) that is neither diagonal nor invariant under geometric changes. As a result, simulations involving structural evolution—such as molecular dynamics—suffer a significant computational overhead.

To address these issues, we have developed a new automated algorithm for model construction in which most parameters are computed rather than fitted, improving model reliability and internal consistency. This method requires tight integration between the model-generation toolkit (\textsc{modelmaker}) and the first-principles code. We use \textsc{siesta}~\cite{Soler-02} as the main DFT engine for this purpose, although the approach can be readily extended to other codes that employ \textsc{wannier90}~\cite{mostofi_cpc08,Pizzi-20} for WF generation.

As described in Sec.~\ref{sec:mani}, we have implemented substantial improvements in \textsc{siesta} to enable independent wannierization of distinct band manifolds.
This section is organized as follows:
(i) In Sec.~\ref{sec:mani}, we describe the selection of band manifolds and the generation of WFs using \textsc{siesta} and \textsc{wannier90};
(ii) Sec.~\ref{sec:symm} explains how symmetry is enforced when computing the integrals entering the second-principles Hamiltonian;
(iii) Sec.~\ref{sec:param} outlines the parameter evaluation strategy based on the previous components;
finally, Sec.~\ref{sec:train_ee} and Sec.~\ref{sec:train_ellat} present the first-principles calculations used to extract the electron-electron and electron-lattice interaction parameters, respectively.

\subsection{Band manifold selection and wannierization procedure}\label{sec:mani}

We define a manifold of $J$ bands as the set of corresponding Bloch-like states that are used to construct $J$ Wannier functions (WFs). A manifold may be either isolated—i.e., energetically separated from other bands throughout the Brillouin zone—or entangled, meaning the target bands overlap and hybridize with others within a broader energy window~\cite{souza_prb01,marzari_rmp12}.

The choice of manifolds included in a second-principles model is critical, as it determines the physical phenomena that can be accurately captured and directly affects the computational cost of simulations. For example, a model for an insulator that includes only the valence (conduction) bands cannot describe optical excitations but may offer greater efficiency when simulating hole (electron) polarons than a model that includes both valence and conduction bands.

By default, \textsc{wannier90}~\cite{mostofi_cpc08,Pizzi-20} performs the wannierization procedure over a single manifold, typically defined by a global energy window. However, in the context of second-principles modeling, it is often desirable to construct WFs from multiple, physically distinct manifolds—for instance, separate treatment of valence and conduction bands, though the approach is not limited to this case.

While one might naively consider applying \textsc{wannier90} independently to each manifold, this presents challenges when matrix elements connecting different manifolds are required. Although Hamiltonian matrix elements in the WF basis, 

\begin{align}
h_{\bm{a}\bm{b}} = \left\langle \chi_{\bm{a}} \right\vert \hat{h} \left\vert \chi_{\bm{b}} \right\rangle,
\end{align}

\noindent vanish for WFs $\chi_{\bm{a}}$ and $\chi_{\bm{b}}$ belonging to different orthogonal manifolds, other matrix elements such as 

\begin{align}\label{eq:WF_position}
\vec{r}_{\bm{a}\bm{b}} = \left\langle \chi_{\bm{a}} \right\vert \vec{r} \left\vert \chi_{\bm{b}} \right\rangle,
\end{align}

\noindent which correspond to dipole transitions, will typically be nonzero and are necessary for simulating processes such as optical excitations.

To address this, we have extended the \textsc{siesta} code~\cite{siesta_jcp20} to enable the simultaneous wannierization of multiple independent manifolds via \textsc{wannier90}. \textsc{siesta} can now generate the input required for \textsc{wannier90} to construct a combined auxiliary manifold containing all previously obtained WFs, enabling the evaluation of inter-manifold matrix elements such as those described in Eq.~\eqref{eq:WF_position}.

In addition, we have identified two practical issues that arise during wannierization. First, small geometric perturbations can lead to discontinuous changes in WF shapes, introducing significant errors in electron-lattice coupling elements calculated via finite differences. Second, the initial projectors used to generate WFs may break symmetry when the spread minimization algorithm is employed, leading to less symmetric and less transferable WFs.
We found that these issues can be mitigated by disabling the spread optimization in \textsc{wannier90}, which results in WFs that are more symmetric and stable under structural changes. We have not found large increases in the number of Hamiltonian terms and Wannier spread. Furthermore, we observed improved localization and symmetry when using \textsc{siesta}’s basis functions as initial projectors, rather than the default hydrogenoid functions typically used by \textsc{wannier90}.

\subsection{Symmetrization of the parameters}\label{sec:symm}

A major limitation in earlier second-principles model construction approaches was the need to explicitly include all symmetry-equivalent configurations in the first-principles training set. On one hand, this led to an unnecessary increase in the number of parameters, as symmetry-related integrals were treated as independent. On the other hand, it required a significantly larger number of first-principles calculations to sample all symmetry-equivalent atomic configurations.

In the present work, we exploit the space group symmetries of the reference atomic geometry (RAG), obtained from the Bilbao Crystallographic Server~\cite{BilbaoCrysI,BilbaoCrysII}, and apply the corresponding symmetry operations $\hat{T}_o$ to the WFs. These transformations yield relations among WFs of the form

\begin{equation}\label{eq:symm_trans}
\hat{T}_o\left\vert\chi_{\bm{a}} \right\rangle = \sum_{\bm{a^\prime}} T^o_{\bm{aa^\prime}} \left\vert \chi_{\bm{a^\prime}} \right\rangle,
\end{equation}

\noindent where $T^o_{\bm{a a}'}$ denotes the matrix elements of the symmetry operation $\hat{T}_o$ in the WF basis.

We apply this transformation to the relevant matrix elements that define the second-principles model, including those entering the total energy expression [Eq.(\ref{eq:totalenergy1updn})], the one-electron Hamiltonian [Eq.(\ref{eq:honeR})], and the expected position operator [Eq.~(\ref{eq:WF_position})]. In the case of the one-electron integrals, and considering that the symmetry operations are defined to leave the Hamiltonian invariant, we obtain

\begin{align}\label{eq:h_symm}
\hat{T}_o\gamma_{\bf ab} & =\hat{T}_o\left\langle\chi_{\bm{a}}\right\vert \hat{h}(n_0) \left\vert \chi_{\bm{b}}\right\rangle \nonumber\\
                  & =  \left\langle\hat{T}_o\chi_{\bm{a}}\right\vert \hat{h}(n_0) \left\vert \hat{T}_o\chi_{\bm{b}}\right\rangle \nonumber\\
                  & = \sum_{\bm{a^\prime b^\prime}} T^{o,T}_{\bm{aa^\prime}} T^o_{\bm{b b^\prime}} \left\langle\chi_{\bm{a^\prime}}\right\vert \hat{h}(n_0) \left\vert \chi_{\bm{b^\prime}}\right\rangle \nonumber\\
                  & = \sum_{\bm{a^\prime b^\prime}} T^{o,T}_{\bm{aa^\prime}} T^o_{\bm{b b^\prime}} \gamma_{\bm{a^\prime b^\prime}}  \nonumber\\
                  & = \gamma_{\bm{ab}},
\end{align}

\noindent where the $T$ superindex indicates transpose. The set of $\gamma_{\bm{a b}}$ elements that are related through nonzero coefficients $T^{o, T}_{\bm{a a}'} T^{o}_{\bm{b b}'}$ are said to form a \textit{one-electron group of symmetry-equivalent parameters}.

Similarly, for the linear electron-lattice coupling constants, we obtain

\begin{align}\label{eq:f_symm}
\hat{T}_o\vec{f}_{\bm{ab},\bm{\lambda}} = \vec{f}^\prime_{\bm{ab},\bm{\lambda}} = \sum_{\bm{a^\prime b^\prime}} T^{o,T}_{\bm{aa^\prime}} T^o_{\bm{b b^\prime}} \vec{f}_{\bm{a^\prime b^\prime},\bm{\lambda^\prime}},
\end{align}

\noindent and for the quadratic electron-lattice couplings

\begin{align}\label{eq:g_symm}
\hat{T}_o\overleftrightarrow{g}_{\bm{ab},\bm{\lambda}\bm{\upsilon}} = \overleftrightarrow{g}^\prime_{\bm{ab},\bm{\lambda\upsilon}} = \sum_{\bm{a^\prime b^\prime}} T^{o,T}_{\bm{aa^\prime}} T^o_{\bm{b b^\prime}} \overleftrightarrow{g}_{\bm{a^\prime b^\prime},\bm{\lambda^\prime\upsilon^\prime}}.
\end{align}

For the electron-electron interaction terms, the symmetry relations become

\begin{align}
   \label{eq:u_symm}
   \hat{T}_oU_{\bm{ab},\bm{a^\prime b^\prime}} & =
        U_{\bm{ab},\bm{a^\prime b^\prime}} 
   \nonumber \\
       & = \sum_{\bm{a^{\prime\prime} b^{\prime\prime}}}
           \sum_{\bm{a^{\prime\prime\prime} b^{\prime\prime\prime}}} 
           T^{o,T}_{\bm{aa^{\prime\prime}}} T^o_{\bm{b b^{\prime\prime}}}
           T^{o,T}_{\bm{a^\prime a^{\prime\prime\prime}}} 
           T^o_{\bm{b^\prime b^{\prime\prime\prime}}} 
           U_{\bm{a^{\prime\prime}b^{\prime\prime}},\bm{a^{\prime\prime\prime} b^{\prime\prime\prime}}},
\end{align}

\noindent and

\begin{align}
   \label{eq:i_symm}
   \hat{T}_o I_{\bm{ab},\bm{a^\prime b^\prime}}  
      & = I_{\bm{ab},\bm{a^\prime b^\prime}} 
   \nonumber \\
      & = \sum_{\bm{a^{\prime\prime} b^{\prime\prime}}}
          \sum_{\bm{a^{\prime\prime\prime} b^{\prime\prime\prime}}} T^{o,T}_{\bm{aa^{\prime\prime}}} T^{o}_{\bm{b b^{\prime\prime}}}
          T^{o,T}_{\bm{a^\prime a^{\prime\prime\prime}}} 
          T^{o}_{\bm{b^\prime b^{\prime\prime\prime}}} 
          I_{\bm{a^{\prime\prime}b^{\prime\prime}},
          \bm{a^{\prime\prime\prime} b^{\prime\prime\prime}}}.
\end{align}

We call Eqs.~(\ref{eq:h_symm})-(\ref{eq:i_symm}) symmetry constraints ($S_{\rm C}$), 
for reasons that will become evident in Sec.~\ref{sec:train_ee}.

In order to obtain $T_{\bm{aa^\prime}}^o$ we simply notice that the WFs obtained from the projection on \textsc{siesta}'s basis orbitals during the wannierization procedure, Sec.~\ref{sec:mani}, need to transform in the same way as these orbitals. 
Comparing the value of the $h_{\bm{ab}}$ elements obtained through \textsc{wannier90} we can check whether the wannierization scheme applied in any practical calculation leads to correct symmetrization.

Another important use of symmetry is the reduction in the number of 
calculations needed to create a {\it training-set} to determine the
value of the various integrals needed in a {\it second-principles}
model. 
In the case of calculations where the geometry changes, like those 
needed to calculate the electron-lattice coupling, see Sec.~\ref{sec:ellat},
a distortion with respect to the RAG, 
realized in a particular geometry of a supercell, is described by the vectors
$\vec{u}_{\bm{\lambda}}$, as described in Eq.~(\ref{eq:atom_pos}).
Calculations that involve the distortions
\begin{align}
\vec{u}_{\bm{\lambda^\prime}} = \hat{T}_o \vec{u}_{\bm{\lambda}}
\end{align}
\noindent do not contain any new information with respect to the carried for $\vec{u}_{\bm{\lambda}}$.
Using the DFT calculation at the geometry given by $\vec{u}_{\bm{\lambda}}$ and applying Eq.~(\ref{eq:h_symm}) it is possible to obtain the values of $h_{\bm{ab}}^{\rm DFT}$ at $\vec{u}_{\bm{\lambda^\prime}}$ by simply operating,
\begin{align}
h_{\bm{a^\prime b^\prime}}^{\rm DFT}(\vec{u}_{\bm{\lambda}^\prime}) & =
\hat{T}_o^{-1} h_{\bm{a^\prime b^\prime}}^{\rm DFT}(\vec{u}_{\bm{\lambda}^\prime})
\nonumber \\
& = \sum_{\bm{a b}} T^{o,-1}_{\bm{a^\prime a}} T^{o,-1}_{\bm{b^\prime b}} h_{\bm{a b}}^{\rm DFT}(\vec{u}_{\bm{\lambda}}).
\end{align}
\noindent This procedure allows us to reduce the number of calculations to obtain
the electron-lattice matrix elements. 
In SrTiO$_3$, the number of single-point calculations necessary to create the electron-lattice model using a 2$\times$2$\times$2 supercell is 57840 without the use of symmetry. 
This is reduced to 15 when using symmetry and an electron-lattice cutoff $\delta r_{\rm el}=2.0$ \AA~ or 209 if $\delta r_{\rm el}=5.6$ \AA, see Sec.~\ref{sec:STO}.
In LiF, using the conventional cell as a supercell, a similar reduction is achieved going from 2353 single-point calculations without symmetry to 21 with symmetry and  $\delta r_{\rm el}=3.0$ \AA.

%{\color{blue} 
%Pablo, coloco aquí una tabla con los cálculos necesarios para calcular las f's y las g's. Por una parte, el número total de cálculos se ve reducido por el cutoff $\delta r_{\rm el}$ y posteriormente se ve reducido por la aplicación de simetría. No sé si quieres comentarlo de forma más amplia o añadir esta tabla con los números más concretos.  
%
%
%\begin{table}[H]
%    \centering
%    \begin{tabular}{c|c|c|c}
%    \hline \hline
%    \multicolumn{4}{c}{$\rm SrTiO_3$ supercell $2\times 2 \times 2$}\\
%    \hline
%          No cutoff + no symmetry & $\delta r_{\rm el}$ (\AA) & Cutoff & Symmetry  \\
%         \hline
%         %
%         \multirow{ 3}{*}{57840} & 2.0 & 612 & 15 \\
%         %
%         & 4.0 & 3456 & 86 \\
%         %
%         & 5.6 & 6588 & 209 \\
%         \hline \hline
%    \end{tabular}
%    \caption{Caption}
%    \label{tab:my_label}
%\end{table}
%
%\begin{table}[H]
%    \centering
%    \begin{tabular}{c|c|c|c}
%    \hline \hline
%    \multicolumn{4}{c}{LiF supercell $1\times 1 \times 1$}\\
%    \hline
%          No cutoff + no symmetry & $\delta r_{\rm el}$ (\AA) & Cutoff & Symmetry  \\
%         \hline
%         %
%         \multirow{ 2}{*}{2353} & 2.1 & 1152 & 9 \\
%         %
%         & 3.0 & 2016 & 21 \\
%         \hline \hline
%    \end{tabular}
%    \caption{Caption}
%    \label{tab:my_label}
%\end{table}
%}

\subsection{Parameter determination}\label{sec:param} 

Second-principles models involve parameters that can either be calculated directly or fitted to data~\cite{wojdel_jpcm13,Escorihuela_prb17,pgf_prb16}. While a direct evaluation offers clearer physical interpretation, its practical implementation may be challenging. For instance, the electron-lattice couplings, $\vec{f}_{\bm{ab},\bm{\lambda}}$ and $\overleftrightarrow{g}_{\bm{ab},\bm{\lambda}\bm{\upsilon}}$, can be obtained via finite-difference methods, as discussed in Sec.~\ref{sec:train_ellat}, by displacing atoms from the RAG and tracking variations in $h_{\bm{ab}}$. These perturbations are straightforward to introduce in standard first-principles frameworks.
In contrast, direct computation of electron-electron interaction terms, $U_{\bm{ab},\bm{a^\prime b^\prime}}$ and $I_{\bm{ab},\bm{a^\prime b^\prime}}$, poses significant difficulties. A finite-difference scheme in this case would require precise manipulation of the density matrix within a DFT code, see Eq.~(\ref{eq:totalenergy1updn}) and Eqs.~(\ref{eq:derhu})--(\ref{eq:derhi}). Although constrained DFT methods~\cite{kaduk_chemrev12} may enable such calculations, their implementation is nontrivial and often code-dependent.
Consequently, we adopt a hybrid strategy: parameters that can be reliably extracted from \textsc{wannier90}, such as $\gamma_{\bm{ab}}$, $\vec{f}_{\bm{ab},\bm{\lambda}}$, $\overleftrightarrow{g}_{\bm{ab},\bm{\lambda}\bm{\upsilon}}$, and $\vec{r}_{\bm{ab}}$, are computed directly, while the electron-electron parameters, $U_{\bm{ab},\bm{a^\prime b^\prime}}$ and $I_{\bm{ab},\bm{a^\prime b^\prime}}$, are obtained via fitting procedures. Efforts to refine this scheme are ongoing.

\subsubsection{One-electron parameters}
\label{sec:train_oneel}

The hopping matrix elements, $\gamma_{\bm{ab}}$, are computed using \textsc{wannier90} from a nonmagnetic ground-state DFT calculation at the RAG. However, enforcing the symmetry condition of Eq.~(\ref{eq:h_symm}) requires high numerical precision in the DFT setup, such as dense $k$-point meshes, which is computationally demanding. An alternative approach, implemented in \textsc{modelmaker}, employs Eq.~(\ref{eq:h_symm}) to symmetrize $\gamma_{\bm{ab}}$ through a weighted average over symmetry-equivalent elements inside a group. The weights are associated to the symmetry matrix components, $T_{\bm{aa^\prime}}^o$ [Eq.~(\ref{eq:symm_trans})] that related one matrix element with another. 

To enforce locality, we introduce a Hamiltonian cutoff distance, $\delta r_{\rm h}$, which eliminates interactions between Wannier functions whose centers exceed this separation. This ensures that the Hamiltonian remains sparse, even if the WFs are not maximally localized, and computationally efficient, enabling linear scaling with system size. 
We also enforce locality for the matrix elements of the position operator, $\vec{r}_{\bm{ab}}$.
In this case we retain all diagonal terms and discard off-diagonal terms ($\bm{a} \neq \bm{b}$) with norm below a given cutoff threshold, $\delta r_{\rm r}$.

\subsubsection{Electron-electron parameters, training set and fitting procedure} \label{sec:train_ee}

Electron-electron parameters are determined by fitting to a training set of DFT-calculated Hamiltonians for a representative ensemble of electronic configurations $\{A\}$. The target is to minimize the deviation between the DFT matrix elements, $h_{\bm{ab}}^\text{DFT}(A)$, and the second-principles model predictions $h_{\bm{ab}}(A, \{p\})$, where $\{p\}$ denotes the set of fitting parameters. Symmetry constraints, Eqs.~(\ref{eq:h_symm})--(\ref{eq:i_symm}), are enforced using Lagrange multipliers during the optimization

\begin{align} 
\Theta = \sum_{\bm{ab}} \sum_A \left[h_{\bm{ab}}(A,\{p\})-h_{\bm{ab}}^\text{DFT}(A)\right]^2 - \sum_{\text C} \lambda_{\text C} S_{\text C}(\{p\}), 
\end{align}

\noindent where $\{\text{C}\}$ denotes the set of independent symmetry constraints, $\lambda_\text{C}$ the associated multipliers, and $S_{\text{C}}(\{p\})$ the corresponding constraint equations. Redundant conditions are excluded to reduce the computational cost of the calculation.

Following Ref.~\cite{pgf_prb16}, the Hamiltonian is partitioned into long-range ($h_{\bm{ab}}^\text{lg}$), associated with electrostatic interactions (charges and dipoles), and short-range ($h_{\bm{ab}}^\text{sh}$) contributions. This separation facilitates a more structured and physically transparent modeling of electron-electron interactions.
Since the fitting procedure only needs to determine the short range part
we can find it by explicitly writing,
\begin{widetext}
\begin{align}
\Theta = \sum_{\bm{ab}} \sum_A \left[h_{\bm{ab}}^\text{sh}(A,\{p\})+h_{\bm{ab}}^\text{lg}(A)-h_{\bm{ab}}^\text{DFT}(A)\right]^2-\sum_{\text C} \lambda_{\text C} S_{\text C}(\{p\}).
\label{eq:goalfunction}
\end{align}
The minimization of the goalfunction $\Theta$ with respect to the parameters,
$\{p\}=\{U_{\bm{ab},\bm{a^\prime b^\prime}}, I_{\bm{ab},\bm{a^\prime b^\prime}}\}$ and the Lagrange multipliers $\lambda_{\rm C}$
leads to the following linear equation system
\begin{align}
\sum_{\bm{ab}} \sum_A h^\text{sh}_{\bm{ab}}(A,\{p\})\frac{\partial h^\text{sh}_{\bm{ab}}}{\partial p_i}(A, \{p\})-\frac{1}{2}\sum_{\text{C}} \lambda_{\text{C}} \frac{\partial S_{\text{C}}}{\partial p_i}(\{p\}) &= \sum_{\bm{ab}} \sum_A \left[h_{\bm{ab}}^\text{DFT}(A)-h_{\bm{ab}}^\text{lg}(A)\right]\frac{\partial h^\text{sh}_{\bm{ab}}}{\partial p_i}(A, \{p\}),\label{eq:ee_fit1}\\
S_{\text{C}}(\{p\})& =0,\label{eq:ee_fit2}
\end{align}
\end{widetext}
\noindent where the derivative of the one electron Hamiltonian, Eq.~(\ref{eq:honeR}), 
with respect to $p_i$ is,

\begin{align}
\frac{\partial h^{\text{sh},s}_{\bm{ab}}}{\partial U_{\bm{ab},\bm{a^\prime b^\prime}}}=&D_{\bm{a}^\prime \bm{b}^\prime}^s+D_{\bm{a}^\prime \bm{b}^\prime}^{-s},\label{eq:derhu}\\
\frac{\partial h^{\text{sh},s}_{\bm{ab}}}{\partial I_{\bm{ab},\bm{a^\prime b^\prime}}}=&-\left(D_{\bm{a}^\prime \bm{b}^\prime}^s-D_{\bm{a}^\prime \bm{b}^\prime}^{-s}\right)\label{eq:derhi}.
\end{align}

\noindent With the equations introduced above, it becomes evident that (i) the parameters $U_{\bm{ab},\bm{a^\prime b^\prime}}$ and $I_{\bm{ab},\bm{a^\prime b^\prime}}$ quantify the sensitivity of the one-electron Hamiltonian, $h_{\bm{ab}}$, to variations in the charge and spin polarization encoded in the density matrix elements $D_{\bm{ab}}$. It is important to emphasize that these quantities do not represent pure electron-electron interactions. Instead, they describe the effective response of the electronic structure to changes in the density matrix, incorporating not only electron-electron interactions but also contributions from the electron kinetic energy and the electron-nuclear potential.
(ii) Accurate evaluation of these parameters requires explicit control over the electronic density—a task that, as previously discussed, is challenging to implement within standard first-principles frameworks. For this reason, $U_{\bm{ab},\bm{a^\prime b^\prime}}$ and $I_{\bm{ab},\bm{a^\prime b^\prime}}$ are obtained by solving the fitting equations, Eqs.~(\ref{eq:ee_fit1})--(\ref{eq:ee_fit2}), using a modified stepwise regression algorithm with forward selection, adapted from the method described in Ref.~\cite{Escorihuela_prb17} to accommodate the variables of the present electronic model.

Having established the fitting framework, the next step is to define a suitable training set for extracting $U_{\bm{ab},\bm{a^\prime b^\prime}}$ and $I_{\bm{ab},\bm{a^\prime b^\prime}}$. Inspection of Eq.~(\ref{eq:derhu}) and Eq.~(\ref{eq:derhi}) reveals that $U_{\bm{ab},\bm{a^\prime b^\prime}}$ can be accessed by varying the total charge of the system, while $I_{\bm{ab},\bm{a^\prime b^\prime}}$ is associated with variations in spin polarization.
Accordingly, the training set includes first-principles simulations in which the number of electrons is systematically modified. This simulates hole or electron doping and probes the influence of occupation changes on the electronic states near the valence and conduction bands. In practice (see Sec.~\ref{sec:results}), these calculations are performed on supercells comprising 30–50 atoms, with total charge varied from $-0.3$ to $+0.3$ $e$ in steps of 0.1 $e$.
To obtain information relevant to $I_{\bm{ab},\bm{a^\prime b^\prime}}$, the same doping configurations are recalculated under the condition that the excess charge is fully spin-polarized. Specifically, the difference in the number of spin-up and spin-down electrons is set equal to the magnitude of the doping charge.
In addition, to accurately capture certain optical excitations, such as excitons, it is advantageous to augment the training set with charge-neutral calculations in which a finite spin polarization is artificially imposed. In insulating systems, such constraints effectively promote electrons from the valence to the conduction band, allowing the model to encode excitonic physics within the fitted electron-electron interaction parameters.

%{\color{red} (PGF: Quitar este párrafo?. JJ: Yo sí lo quitaría. Sabemos que tenemos un problema ahí, pero no creo que interfiera con lo que estamos tratando en este artículo. Es algo que puede quedar para más adelante. Y no creo que nadie se de cuenta del mismo en estos momentos.)
%These calculations are particularly important to describe polarons and excitons in
%{\it second-principles}.
%
%We have noticed that performing these calculations in the RAG does not lead to
%an adequate performance of the models in this respect.
%
%The main reason is that simple doping leads to a fully delocalized state where
%the effect on the WFs mostly involved in the polaronic/excitonic state are not 
%influenced. 
% 
%A possibility to reduce this problem is to dope in a distorted geometry, 
%although this is somewhat removed from the goal of achieving a fully automated
%parameterization scheme. 
%
%We continue working to solve these problems.
%}

An additional critical consideration is the selection of which specific $U_{\bm{ab},\bm{a^\prime b^\prime}}$ and $I_{\bm{ab},\bm{a^\prime b^\prime}}$ matrix elements should be included in the model. As these quantities depend on the spatial positions of four Wannier functions (WFs), their total number can become prohibitively large. This challenge is compounded by the difficulty of perturbing individual elements of the density matrix using the {\it training set} methodology outlined above.
To address this, we adopt two complementary criteria to determine which four-center integrals are retained in the fit. First, we impose a spatial cutoff, $\delta r_{\mathrm{ee}}$, that limits the maximum distance between the centroids of any pair of WFs involved in the integral. Second, we evaluate the contribution of groups of Hamiltonian elements to the goalfunction, Eq.~(\ref{eq:goalfunction}), when the second-principles model includes only the $\gamma_{\bm{ab}}$ parameters computed at the reference (not relaxed) atomic geometry (RAG) for the reference electronic density (RED). If the cumulative contribution of a group exceeds a threshold $\delta \Theta$ (in units of eV$^2$) across the training set, the corresponding $U_{\bm{ab},\bm{a^\prime b^\prime}}$ and $I_{\bm{ab},\bm{a^\prime b^\prime}}$ integrals are included in the model. Additional examples and parameter values are provided in Sec.\ref{sec:results}.

\subsubsection{Electron-lattice terms}\label{sec:train_ellat}

Electron-lattice coupling terms are computed using finite-difference approximations applied to the one-electron Hamiltonian obtained from density functional theory (DFT). Specifically, the $x$-component of the linear electron-lattice coupling vector $\vec{f}_{\bm{ab},\bm{\lambda}}^{\hspace{0.5mm}\mathrm{fd}}$ is estimated using the central difference formula

\begin{align}
f_{\bm{ab},\bm{\lambda}}^{{\rm fd}, x} \approx - \frac{h_{\bm{ab},\bm{\lambda}+x}^{\text{DFT}}-h_{\bm{ab},\bm{\lambda}-x}^{\text{DFT}}}{2\delta_x}
\label{eq:fd_linear}
\end{align}

\noindent where $h_{\bm{ab},\bm{\lambda}\pm x}^{\text{DFT}}$ denotes the $\bm{ab}$ matrix element of the Hamiltonian in the Wannier basis, with atom $\bm{\lambda}$ displaced by $\pm\delta_x$ along the direction $x$, that represents a generic displacement along any of the cartesian ($x$, $y$ or $z$) directions.
The quadratic electron-lattice coupling matrix is similarly obtained as 

\begin{equation}
\resizebox{0.48\textwidth}{!}{$
\begin{aligned}
g_{\bm{ab},\bm{\lambda\upsilon}}^{xy} & \approx  \\ 
& \frac{h_{\bm{ab},\bm{\lambda}+x\hspace{0.5mm}\bm{\upsilon}+y}^{\text{DFT}}-h_{\bm{ab},\bm{\lambda}+x\hspace{0.5mm}\bm{\upsilon}-y}^{\text{DFT}}
 -h_{\bm{ab},\bm{\lambda}-x\hspace{0.5mm}\bm{\upsilon}+y}^{\text{DFT}}+h_{\bm{ab},\bm{\lambda}-x\hspace{0.5mm}\bm{\upsilon}-y}^{\text{DFT}}}
{4\delta_x \delta_y},\label{eq:fd_quad}
\end{aligned}$}
\end{equation}

\noindent where the notation $h_{\bm{ab},\bm{\lambda}\pm x,\bm{\upsilon}\pm y}^{\text{DFT}}$ refers to matrix elements computed with atoms $\bm{\lambda}$ and $\bm{\upsilon}$ displaced along $x$ and $y$, respectively.
All necessary displacements are performed within the same supercell used for the electron-electron interaction calculations. However, direct application of Eq.(\ref{eq:fd_linear}) and Eq.~(\ref{eq:fd_quad}) would require an impractically large number of DFT calculations. To reduce computational cost, we employ two strategies. First, the symmetry operations of the RAG space group are applied to identify symmetry-equivalent atomic configurations, which are not recalculated (see Sec.~\ref{sec:symm}). Second, we impose a distance cutoff $\delta r_{\mathrm{el}}$ that restricts the maximum allowed separation between displaced atoms measured in the RAG

\begin{align} \delta r_{\rm el} > \left\vert \vec{R}_{\Upsilon} - \vec{R}_{\Lambda} + \vec{\tau}_{\upsilon} - \vec{\tau}_{\lambda} \right\vert, \end{align}

\noindent under the assumption that displacements of distant atoms are sufficiently uncorrelated as far as the corresponding electron-lattice couplings are concerned.
Despite these reductions, the number of computed linear and quadratic constants remains large, with many contributing negligibly to $h_{\bm{ab}}$. Therefore, we retain only those vectors $\vec{f}_{\bm{ab},\bm{\lambda}}$ and matrices $\overleftrightarrow{g}_{\bm{ab},\bm{\lambda\upsilon}}$ for which at least one component exceeds predefined cutoffs $\delta f$ and $\delta g$, expressed in eV/\AA~and eV/\AA$^2$, respectively. In practice, values around $0.1$ eV/\AA~and $0.1$ eV/\AA$^2$ yield satisfactory results (see Sec.~\ref{sec:results}).

The treatment of metallic systems presents additional challenges, as the ground-state density matrix varies with atomic geometry. In such cases, $U_{\bm{ab},\bm{a^\prime b^\prime}}$ (and, optionally, $I_{\bm{ab},\bm{a^\prime b^\prime}}$) are first determined, and the corresponding contributions are subtracted from $h_{\bm{ab}}^{\text{DFT}}$ prior to evaluating $\vec{f}_{\bm{ab},\bm{\lambda}}$ and $\overleftrightarrow{g}_{\bm{ab},\bm{\lambda\upsilon}}$, thus avoiding double-counting of geometry- and density-driven effects.

\subsection{Model validation}
\label{sec:verify}

To evaluate the performance of our models, we employ a goalfunction similar to $\Theta$, defined in Eq.~(\ref{eq:goalfunction}), applied to a {\it test-set} of density functional theory (DFT) calculations, denoted as $\{A\}$. These test sets consist of configurations not included in the training set, selected to probe the model's ability to generalize to configurations representative of physically relevant distortions.
Specifically, test geometries are generated by displacing each atom randomly within a cube of side length $2d$, centered at the RAG position, where $d$ is a parameter controlling the displacement amplitude. In Sec.~\ref{sec:STO} and Sec.~\ref{sec:lif}, the error function $\Theta$ is evaluated by averaging over ten such randomly distorted geometries for each value of $d$.
Additionally, when available, we assess model performance on alternative stable phases not included in the training. For example, in the case of SrTiO$_3$, we include both the tetragonal I4/mcm phase and a ferroelectric phase stabilized under uniaxial strain.
In other systems, we explore the formation of polarons by introducing suitable lattice distortions and comparing the resulting electronic and structural responses—governed by both electron-lattice and electron-electron interactions—with those obtained from DFT calculations.

% ****************************************************************

\section{Computational details}\label{sec:computational}

First-principles calculations were performed using \textsc{siesta}~\cite{Soler-02, Garcia-20}, which is based on the standard Kohn-Sham self-consistent formulation of DFT and employs a numerical atomic orbital (NAO) basis set.

% Functional ------------------------
In order to test the method we employed different exchange-correlation functionals for each of the studied systems.
For $\rm SrTiO_3$ the exchange-correlation functional was approximated within the local density approximation (LDA), adopting the usual Ceperley and Alder~\cite{Ceperley-80, Perdew-81} parametrization.  
In the case of $\rm LiF$, the exchange and correlation was treated using the Perdew-Burke-Ernzerhof (PBE) generalized gradient approximation~\cite{Perdew-96}. 
%
%Both exchange-correlation functionals were implemented as provided in the {\sc libxc} library~\cite{Marques2012libxc,Lehtola2018recent}.

% Basis set ------------------------
For $\rm SrTiO_3$, the parameters that define the shape and spatial extent of the basis functions were determined by computing the eigenfunctions of isolated atoms confined within the soft-confinement potential proposed in Ref.~\cite{Junquera-01}.
We used a single-$\zeta$ basis set for the semicore states of Ti and Sr,  and double-$\zeta$ plus polarization orbitals for the valence states of all the atoms. For Sr an extra shell of 4$d$ orbitals was added.
All the parameters that define the shape and the range of the basis functions for Sr, Ti, and O were obtained by a variational optimization following the recipe given in Ref.~\cite{Junquera-01}.
%A single-$\zeta$ basis set was employed for the semicore states of Sr ($4s$, $4p$) and Ti ($3s$, $3p$). For the valence states of Sr, Ti, and O ($5s$, $4d$ for Sr; $4s$, $3d$ for Ti; and $2s$, $2p$ for O), basis sets of varying sizes were utilized, ranging from single-$\zeta$ for Sr$^{2+}$ to double-$\zeta$ for Ti$^{4+}$ and O$^{2-}$.
%
%To improve angular flexibility in the description of the electronic structure, higher angular momentum polarization orbitals (which are unoccupied in the free atom) were incorporated. Specifically, for Sr, the $4d$ orbitals were treated with a single-$\zeta$ basis. For Ti, the $4p$ orbitals were included with a single radial function per polarization shell. Additionally, for O, a single-$\zeta$ $3d$ orbital was introduced to enhance the description of polarization effects.

For the LiF system, a double-$\zeta$ basis set was employed for the $2s$ and $2p$ orbitals of Li, while a double-$\zeta$ polarized basis set was selected for F. 
In both cases, the spatial extent of the orbitals was defined using the default parameters.  

% Pseudopotentials ----------------
The core electrons were replaced by {\it ab initio} norm-conserving pseudopotentials~\cite{Hamann-79}. The pseudopotentials have been generated  using the Troullier-Martins scheme~\cite{Troullier-91} in the Kleinman-Bylander fully non-local separable representation~\cite{Kleinman-82}. 
%They are expressed in the ``Froyen'' (psf) format,  available at the Virtual Vault for Pseudopotentials site. 
%

% K-point mesh --------------------
The electronic density, Hartree potential, and exchange-correlation potential, along with the corresponding matrix elements between the basis orbitals, were computed on a uniform real-space grid~\cite{Soler-02}. To accurately represent the charge density in both systems, an equivalent plane-wave cutoff of 600 Ry was employed.  
For the reciprocal-space integration, considering the conventional cell geometries of the systems, the integrals were ensured to be well converged. In all cases, a Monkhorst-Pack mesh of equivalent quality was utilized, specifically a \(9 \times 9 \times 9\) grid for the $\rm SrTiO_3$ system and a \(8 \times 8 \times 8\) grid for the LiF system ~\cite{Monkhorst-76}.
The RAG was relaxed for both systems until the maximum component of the stress tensor was below  0.003 eV/$\rm{\AA}^{3}$, atomic forces being zero by symmetry.

% *****************************
%           WANNIER90
% *****************************

The second-principles code \textsc{scale-up} and, by extension, the program generating the models, \textsc{modelmaker}, employ a basis of Wannier functions~\cite{Marzari-97, Marzari-12}, as this provides a compact and short-range basis that is ideal for large-scale simulation methods. 
These functions are constructed by projecting Bloch states onto \textsc{siesta}'s atomic basis orbitals, which serve as initial guess functions, as implemented in the \textsc{wannier90} code~\cite{Pizzi-20,mostofi_cpc08}.
To avoid symmetry problems arising from the minimization of the spread the WFs are directly taken from these projections, after an orthonormalization step. 
A uniform $4 \times 4 \times 4$ reciprocal-space mesh is used to compute the Wannier functions and evaluate relevant quantities in real space.  

\textsc{modelmaker} is a Python-based software package developed as an integrated framework to generate second-principles models from first-principles data. Although currently interfaced with \textsc{siesta}, the code can be readily adapted to other first-principles codes compatible with \textsc{wannier90}.
The generated models include detailed information on the electronic band structure, electron-lattice coupling, and electron-electron interactions. These data are stored in a human-readable \texttt{.xml} format for direct use by \textsc{scale-up}~\cite{pgf_prb16}.

To build a model, \textsc{modelmaker} requires as input a template of the \textsc{siesta} input file corresponding to the RAG, which is used to construct the necessary auxiliary inputs. Additionally, the symmetry operations of the RAG space group—obtained via the Bilbao Crystallographic Server~\cite{BilbaoCrysI}—must be provided. Finally, the calculation of electrostatic corrections requires the Born effective charges of all atoms in the unit cell and the high-frequency dielectric tensor, $\varepsilon_\infty$.

\section{Results}
\label{sec:results}

To demonstrate the application of the methodology described above, we consider two representative systems.

The first one is the semicovalent transition-metal perovskite SrTiO$_{3}$, a material that exhibits strong electron-lattice coupling arising from two key structural instabilities: the incipient ferroelectric distortion~\cite{zhong_prl96, li_prb06}, and the antiferrodistortive octahedral rotations associated with the cubic-to-tetragonal phase transition observed experimentally.

The second example is LiF, a wide-band-gap insulator with the rock-salt structure. In this system, the strong electron-hole interaction is known to lead to prominent optical effects, including the formation of tightly bound excitons~\cite{block_ssc78, song_excitonbook, williams_jpcs90}.

In both cases, we focus our analysis on the procedure used to determine the model parameters outlined in Sec.~\ref{sec:param}. This includes the selection of the relevant Wannier manifolds, the choice of the Hamiltonian interaction cutoff distance, $\delta r_{\rm h}$, the electron-electron interaction cutoff, $\delta r_{\rm ee}$, and the corresponding goal-function convergence threshold, $\delta \Theta$. For the electron-lattice interaction, we additionally define a spatial cutoff, $\delta r_{\rm el}$, as well as first- and second-order interaction cutoffs, denoted $\delta f$ and $\delta g$, respectively.

\subsection{SrTiO$_3$}
\label{sec:STO}

The first system we consider is SrTiO$_{3}$, a widely studied transition-metal perovskite oxide that serves as a common substrate for the epitaxial growth of other perovskites, such as ferroelectric PbTiO$_{3}$~\cite{sanchez_chemsocrev14, kawasaki_appsurfsci96}.
SrTiO$_{3}$ is also a key component in the design of complex oxide superlattices exhibiting intricate polarization patterns~\cite{das_nature19}, has been associated with emergent metallic behavior at interfaces~\cite{ohtomo_nat04}, and is commonly used (when doped) as a bottom electrode in perovskite-based systems.

The RAG adopted for this study corresponds to the high-symmetry cubic paraelectric phase (space group Pm$\bar{3}$m), which contains five atoms per unit cell. Using the \textsc{siesta} code within the LDA, the equilibrium lattice parameter is found to be 3.874 \AA.

\begin{figure}[h]  
    \includegraphics[width=\linewidth]{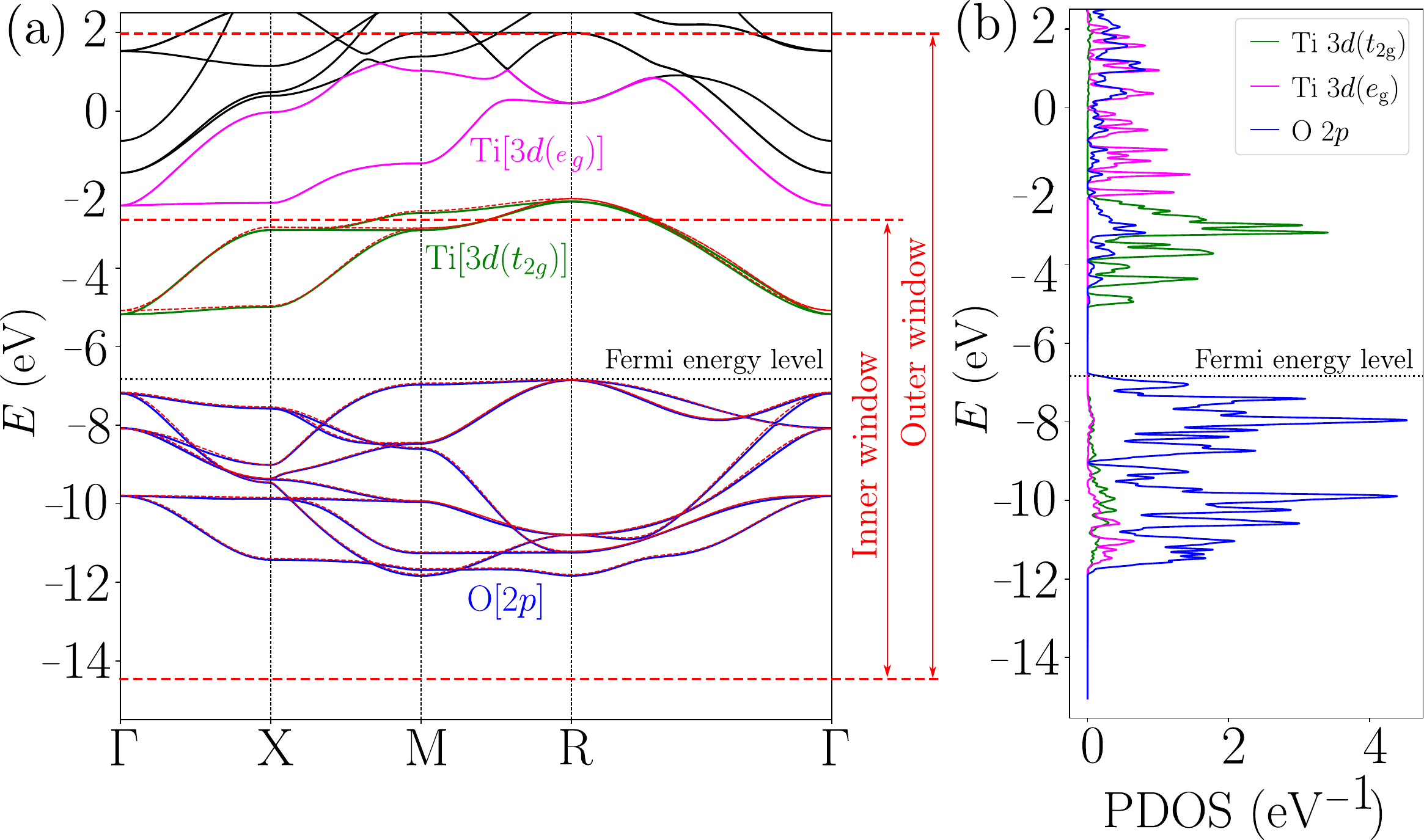}
    \caption{\justifying (a) Electronic band structure of SrTiO$_3$ in the cubic, centrosymmetric reference atomic geometry (RAG). Solid lines represent DFT bands while red dashed curves represent Wannier interpolation. Solid lines in color are associated with DOS projections (see (b) panel). In particular, the valence band manifold with predominant O~(2$p$) character is shown in blue, while the Ti~(3$d$) $t_{\rm 2g}$ and $e_{\rm g}$ conduction band manifolds are shown in green and magenta, respectively. The global energy windows used in the wannierization procedure are indicated by dashed red lines. (b) Corresponding projected density of states (PDOS).
    }
    \label{fig:sto-bands-windows}
\end{figure}

The electronic structure of SrTiO$_{3}$ is characterized by an insulating gap separating the O(2$p$)-dominated valence bands and Ti(3$d$)-dominated conduction bands, as shown in Fig.~\ref{fig:sto-bands-windows}. Note that the bottom of the conduction band consists primarily of Ti(3$d_{xy}$), Ti(3$d_{xz}$), and Ti(3$d_{yz}$) orbitals forming a $t_{\rm 2g}$ manifold (in green), as expected for isolated TiO$_{6}$ octahedra. While the experimental indirect gap is approximately 3.25 eV~\cite{vanbenthem_jap01}, the DFT-LDA result yields a smaller value near 1.7 eV (see Fig.~\ref{fig:sto-bands-windows}).
To construct a second-principles model capable of capturing essential physical phenomena in SrTiO$_{3}$, including polaron formation, we build a Hamiltonian incorporating Wannier functions that describe both the valence band and the $t_{\rm 2g}$ conduction manifold.

Figure~\ref{fig:sto-bands-windows} illustrates the inner and outer energy windows used in the wannierization process. The inner window encompasses the entire valence band and most of the $t_{\rm 2g}$ conduction states, but avoids higher-energy regions where Ti(3$d_{z^2}$) and Ti(3$d_{x^2-y^2}$) orbitals (i.e., the $e_{\rm g}$ manifold) contribute. As described in Sec.~\ref{sec:mani}, valence and conduction bands are wannierized separately using \textsc{siesta}, and subsequently combined to evaluate matrix elements of the position operator $\vec{r}_{\bm{ab}}$.

\begin{figure}[h]
    \centering
    \includegraphics[width=0.9\columnwidth]{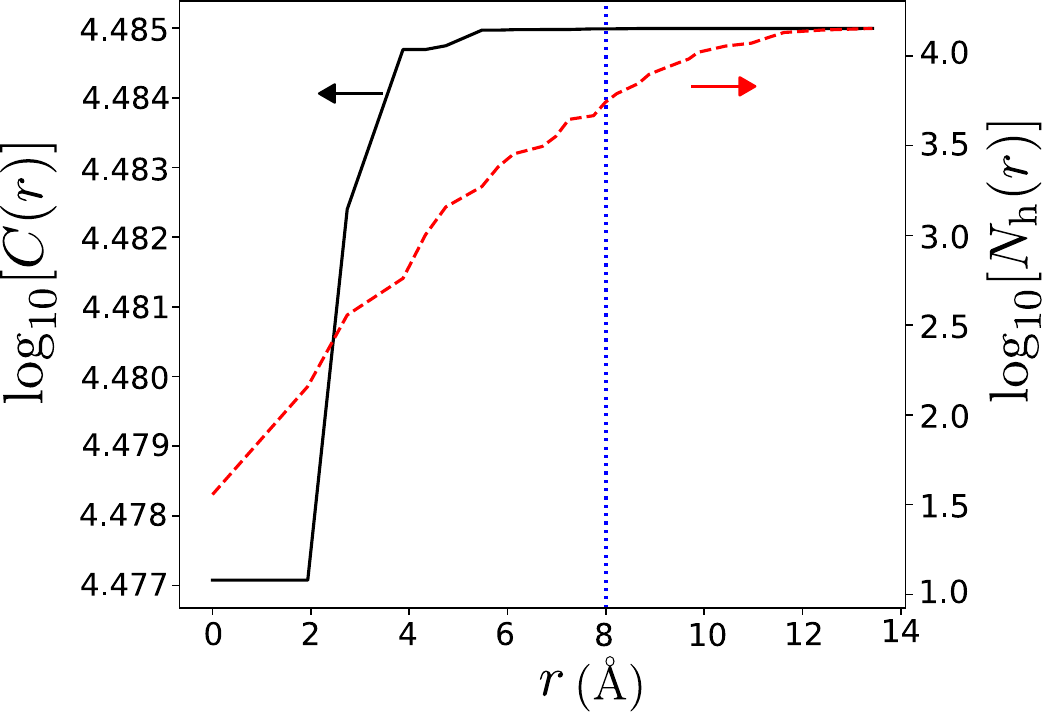} 
    \caption{\justifying (Color online) Black solid line (left axis): Decimal logarithm of the cumulative function $C(r)$, quantifying the contribution of one-electron Hamiltonian elements as a function of the inter-Wannier distance $r$. Red dashed line (right axis): Decimal logarithm of the number of Hamiltonian matrix elements with centroid separation less than $r$.
    The vertical dotted line marks the chosen one-electron Hamiltonian cutoff distance, $\delta r_{\rm h}$ = 8.0 \AA, used in the SrTiO$_{3}$ model to reproduce the band structure shown with dashed red lines in Fig.~\ref{fig:sto-bands-windows}(a).}
    \label{fig:onele-sto}
\end{figure}

To determine an appropriate value for the one-electron interaction cutoff distance $\delta r_{\rm h}$, we compute the function

\begin{align}\label{eq:h_fun}
C(r)=\sum_{\vert \vec{r}_{\bm{b}}-\vec{r}_{\bm{a}}\vert < r} h_{\bm{ab}}^2(r),  
\end{align}

\noindent which sums the squared Hamiltonian matrix elements between pairs of Wannier functions $\bm{a}$ and $\bm{b}$ with centroids separated by less than distance $r$.
As shown in Fig.~\ref{fig:onele-sto}, $C(r)$ remains nearly constant for distances up to 2.1 \AA, corresponding to the Ti–O bond length, and then rises sharply, saturating at around 6.0 \AA\ and reaching a near-plateau by approximately 8.0 \AA.
While a cutoff of $\delta r_{\rm h}$ = 6.0 \AA\ may capture the majority of interactions, we find that using this value we still have small errors near the $\Gamma$-point in the valence band. Therefore, to ensure an accurate representation across the entire Brillouin zone, we select $\delta r_{\rm h}$ = 8.0 \AA. The bands corresponding to the RAG utilizing this cutoff are depicted as dashed red lines in Fig.~\ref{fig:onele-sto}. It is important to note that increasing $\delta r_{\rm h}$ substantially enlarges the number of nonzero Hamiltonian matrix elements (see red curve in Fig.~\ref{fig:onele-sto}), which directly impacts the computational cost of subsequent model simulations.

Having established an accurate representation of the RED in the preceding steps, we now proceed to determine the electron-lattice coupling terms—specifically, the linear couplings $\vec{f}_{\bm{ab},\bm{\lambda}}$ and the quadratic couplings $ \overleftrightarrow{g}_{\bm{a}\bm{b},\bm{\lambda}\bm{\upsilon}}$.
We begin by defining the electron-lattice interaction cutoff distance, $\delta r_{\rm el}$, which sets the maximum separation between atoms participating in the quadratic electron-lattice coupling (as determined by using the atomic positions in the RAG). It is important to emphasize that increasing $\delta r_{\rm el}$ significantly impacts computational cost, both in terms of the number of DFT calculations required to generate the training set and the resulting size of the second-principles model, which in turn affects the computational cost of the corresponding simulations.

\begin{figure*}
    \centering
    \includegraphics[width=0.8\textwidth]{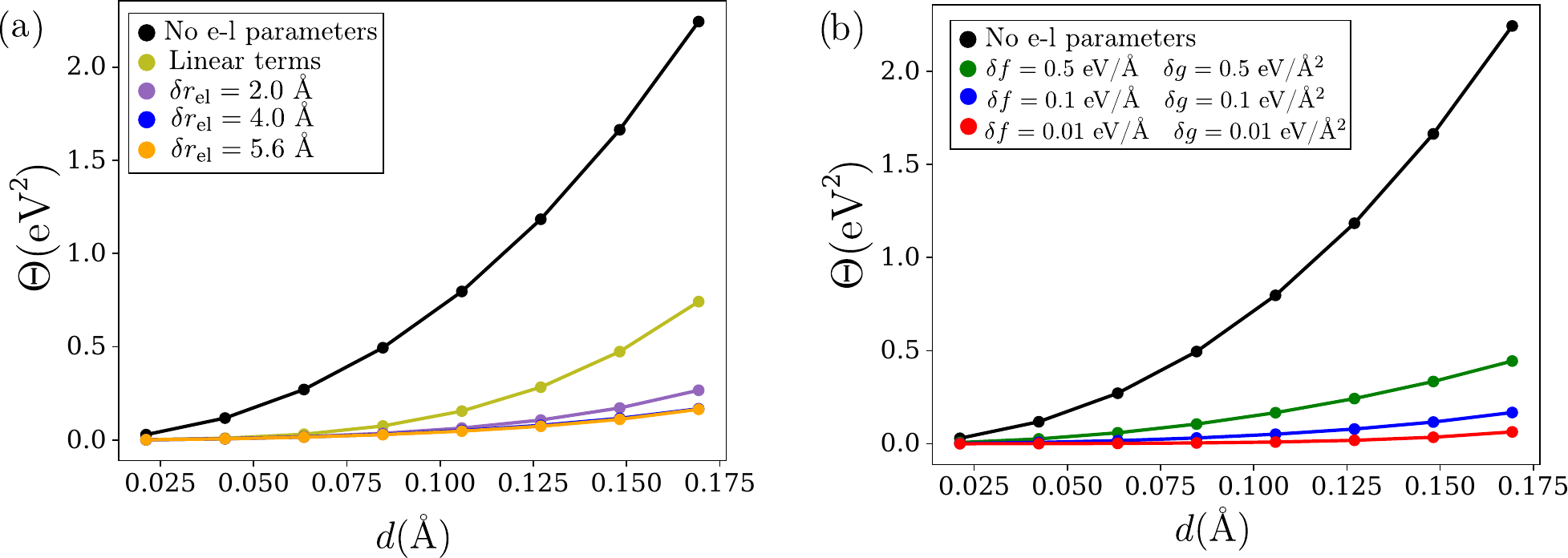}
    \caption{\justifying
    (a) Goalfunction $\Theta$ for SrTiO$_3$ measuring the average error per calculation in {test set} consisting of ten configurations with increasing random displacements $d$ from the RAG. The cutoff values for the electron-lattice force constants are fixed at $\delta f = 0.1$ eV/\AA\ and $\delta g = 0.1$ eV/\AA$^2$.  The black curve corresponds to a model without electron-lattice coupling; the olive green curve includes only linear terms. The purple, blue and orange curves include both linear and quadratic terms, with electron-lattice cutoffs \(\delta r_{\rm el} = 2.0\) \AA{}, \(\delta r_{\rm el} = 4.0\) \AA{} and \(\delta r_{\rm el} = 5.6\) \AA{}, respectively. The blue curve is indistinguishable from the orange one. (b) Effect of varying the cutoff thresholds \(\delta f\) and \(\delta g\) on \(\Theta\), for fixed \(\delta r_{\rm el} = 4.0\) \AA. The green, blue, and red curves correspond to (\(\delta f\), \(\delta g\)) values of (0.5 eV/\AA{}, 0.5 eV/\AA\(^2\)), (0.1 eV/\AA{}, 0.1 eV/\AA\(^2\)), and (0.01 eV/\AA{}, 0.01 eV/\AA\(^2\)), respectively.}
    \label{fig:sto-goalfunctions}     
\end{figure*}

To evaluate the impact of different $\delta r_{\rm el}$ values, we construct a test set composed of 10 randomly distorted geometries, as described in Sec.~\ref{sec:verify}, characterized by an atomic displacement parameter $d$. The resulting orbital energies are compared against their DFT counterparts, and the average error per calculation in testing set is plotted in Fig.~\ref{fig:sto-goalfunctions}(a) as a function of $d$.
The black curve corresponds to a model without any electron-lattice corrections, while the olive green curve includes only linear electron-lattice terms. The remaining curves (purple, blue, and orange) incorporate quadratic corrections for increasing values of $\delta r_{\rm el}$: 2.0 \AA\, which includes only nearest-neighbor Ti–O interactions; 4.0 \AA\, capturing interactions between all the atoms spanning a full five-atom unit cell; and 5.6 \AA\, which incorporates longer-range contributions beyond a single unit cell.

As shown in Fig.~\ref{fig:sto-goalfunctions}(a), the model with $\delta r_{\rm el}$ = 2.0~\AA\ exhibits a more rapid accumulation of error compared to the models with $\delta r_{\rm el}$ = 4.0~\AA\ and 5.6~\AA\, which yield nearly indistinguishable results. This behavior underscores the importance of including quadratic oxygen–oxygen interactions within a single TiO$_{6}$ octahedron in order to accurately capture the modifications to the Hamiltonian induced by the relatively strong electron-lattice coupling in SrTiO$_{3}$.
To reduce the number of electron-lattice coupling terms and thereby simplify the model, we introduce cutoff thresholds $\delta f$ and $\delta g$ for the inclusion of linear and quadratic coupling constants, respectively. The effect of these thresholds on model accuracy is presented in Fig.~\ref{fig:sto-goalfunctions}(b). For large thresholds ($\delta f$ = 0.5 eV/\AA, $\delta g$ = 0.5 eV/\AA$^{2}$, represented in green), the error increases rapidly with the displacement $d$. Reducing both $\delta f$ and $\delta g$
by a factor of five ($\delta f$ = 0.1 eV/\AA, $\delta g$ = 0.1 eV/\AA$^{2}$, shown in blue) leads to a fourfold reduction in error for $d$ = 0.17~\AA. 
These errors can be further reduced until we reach thresholds that are one order of magnitude smaller, 
 $\delta f$ = 0.01 eV/\AA, $\delta g$ = 0.01 eV/\AA$^{2}$, represented in red in Fig.~\ref{fig:sto-goalfunctions}(b), where we seem to find the limit for the quadratic coupling approximation  employed here.

\begin{figure*}[t!]
    \centering
     \includegraphics[width=0.9\linewidth]{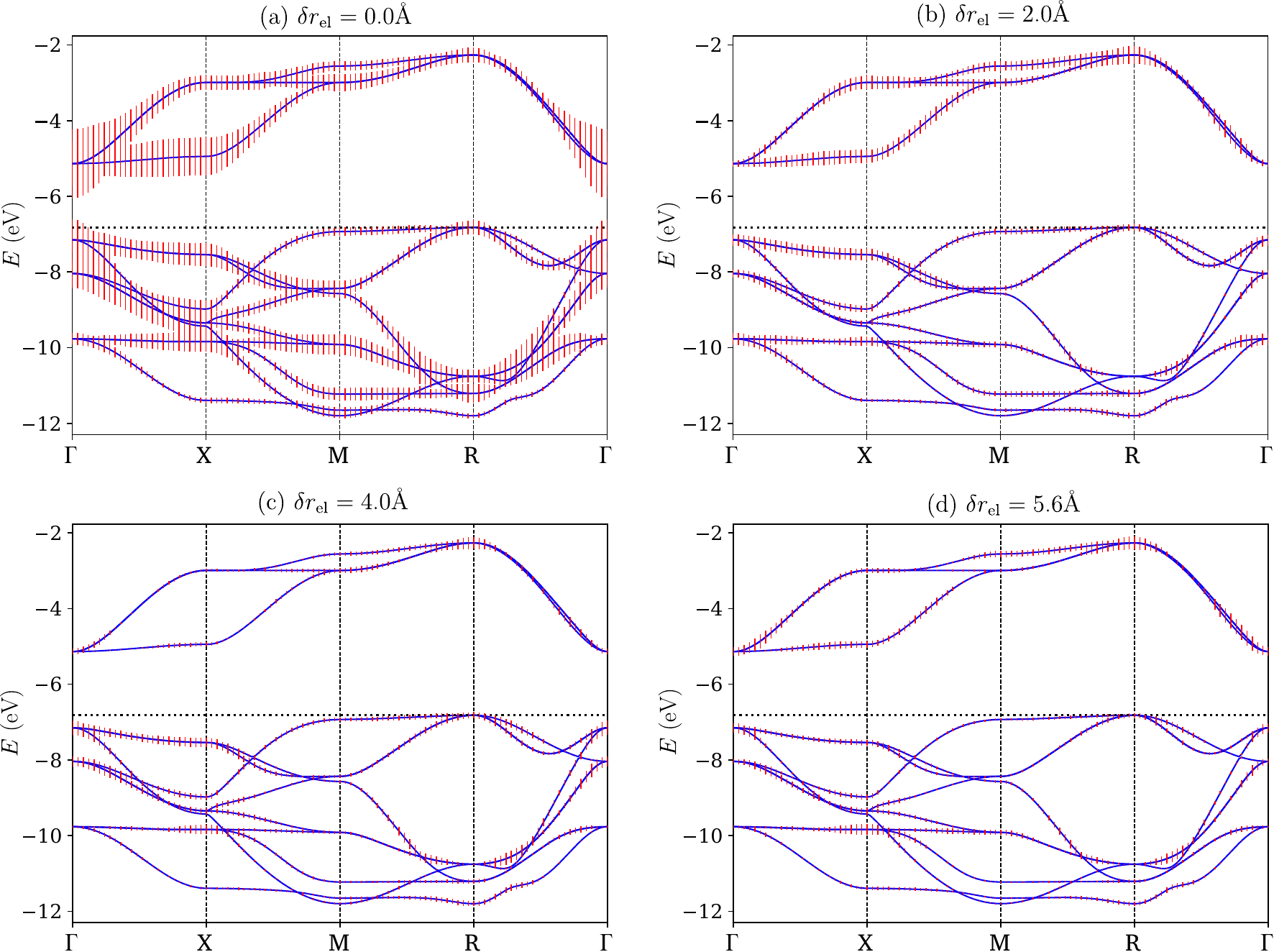}
     \caption{\justifying (Color online) Electronic band structures of SrTiO$_3$ with error bars (red) indicating the statistical deviation of second-principles simulations from DFT results over a test set of ten geometries characterized by $d=0.17$~\AA. Panel (a) corresponds to a model without electron-lattice interaction. Panels (b)–(d) include both linear and quadratic electron-lattice coupling with increasing interaction range characterized by $\delta r_{\rm el} = 2.0$, 4.0, and 5.6~\AA, respectively. All models employ threshold parameters $\delta f = 0.1$ eV/\AA\ and $\delta g = 0.1$ eV/\AA$^2$.}
     \label{fig:sto_bars_varing_r_el}
\end{figure*}

To assess how these errors affect the electronic structure, Fig.~\ref{fig:sto_bars_varing_r_el} presents band structures for SrTiO$_{3}$ including \emph{error bars}. These error bars represent the statistical deviation of the second-principles model predictions from the DFT results for a test set with $d= 0.17$~\AA.
Similar trends are observed for smaller displacements, although in those cases the errors are much smaller than the one presented in Fig.~\ref{fig:sto_bars_varing_r_el}. 
To examine the influence of the electron-lattice cutoff $\delta r_{\rm el}$, we compare the results of the test set for different cutoff values. Figure~\ref{fig:sto_bars_varing_r_el} shows, as a reference the DFT bands for the RAG and each panel includes error bars associated to different situations. In Figure~\ref{fig:sto_bars_varing_r_el}(a) we do not include any electron-lattice coupling so the error bars here quantify the absolute modification of the band energy by the atomic displacements considered. In contrast Figs.~\ref{fig:sto_bars_varing_r_el}(b)–\ref{fig:sto_bars_varing_r_el}(d) present results for $\delta r_{\rm el}$ = 2.0, 4.0, and 5.6~\AA\, respectively, where the error bars quantify the discrepancy between the predictions of our model and the DFT results.
A key result of this analysis is the essential role of electron-lattice coupling parameters in accurately reproducing the electronic band structure across different atomic configurations.
Both the linear electron-lattice contributions and the quadratic terms associated with first-neighbor Ti–O interactions are found to be critical.
This is demonstrated by the substantial reduction in the error bars of the electronic structure when progressing from a model without any electron-lattice coupling [Fig.~\ref{fig:sto_bars_varing_r_el}(a)] to those incorporating linear and quadratic terms for increasing values of $\delta r_{\rm el}$ [Figs.~\ref{fig:sto_bars_varing_r_el}(b)–\ref{fig:sto_bars_varing_r_el}(d)].
Although the inclusion of first-neighbor Ti–O interactions significantly improves the agreement with the reference calculations, the remaining discrepancies in Fig.~\ref{fig:sto_bars_varing_r_el}(b) indicate that this level of approximation is insufficient for full accuracy.

\begin{figure*}[t]
    \centering
     \includegraphics[width=0.9\linewidth]{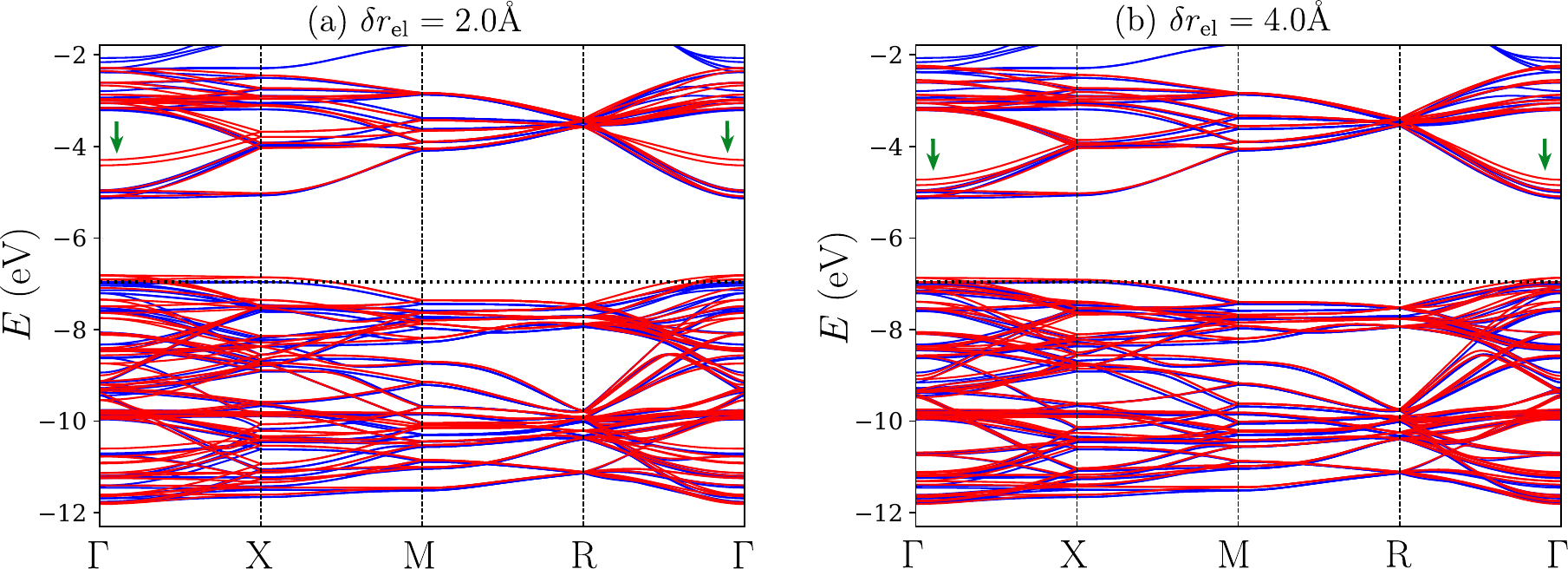}
     \caption{\justifying (Color online) Electronic band structure of SrTiO$_{3}$ in the relaxed antiferrodistortive phase, characterized by in-plane octahedral rotations, with lattice parameters fixed to their cubic values. First-principles (DFT) results are shown in blue, while second-principles results appear in red. Panels (a) and (b) correspond to models constructed with electron-lattice interaction cutoff distances $\delta r_{\rm el} = 2.0$ and 4.0~\AA, respectively. Green arrows highlight band features whose energies exhibit strong sensitivity to $\delta r_{\rm el}$, underscoring the importance of including extended quadratic interactions.}
     \label{fig:sto_bands_rotZ}
\end{figure*}

A more detailed comparison between Figs.~\ref{fig:sto_bars_varing_r_el}(b) and \ref{fig:sto_bars_varing_r_el}(c) underscores the necessity of incorporating longer-range electron-lattice interactions.
In particular, extending the interaction range to include atoms across an entire five-atom unit cell yields a marked reduction in error bars, highlighting the importance of non-local effects.
Notably, interactions between oxygen atoms and their periodic images within a single unit cell are especially relevant, as they are closely tied to octahedral rotation modes.
This becomes particularly evident when analyzing distorted structures such as the antiferrodistortive phase, where the choice of $\delta r_{\rm el}$ relative to the lattice constant has a decisive impact on the quality of the model, as shown in Fig.~\ref{fig:sto_bands_rotZ}.
Specifically, using a cutoff $\delta r_{\rm el} = 2.0$~\AA\ results in significant deviations in the band dispersion along the $\Gamma$–$X$ path, which are largely corrected by including longer-range oxygen–oxygen interactions.
The remaining discrepancies will be analyzed in further analysis.

At this stage, by increasing the electron-lattice cutoff distance, we find that quadratic interatomic interactions beyond $\delta r_{\rm el} = 4$~\AA\ exert a negligible influence on the electronic structure. This conclusion is supported by the results in Fig.~\ref{fig:sto_bars_varing_r_el}(d), as well as by the close overlap between the orange and blue curves in Fig.~\ref{fig:sto-goalfunctions}(a). Thus, a cutoff of $\delta r_{\rm el} = 4.0$~\AA\ offers an optimal compromise between computational efficiency and accuracy.

In terms of model precision, it is notable that the deviations in the valence and conduction bands remain small even for relatively large atomic displacements of up to 0.17~\AA\, as illustrated in Fig.~\ref{fig:sto_bars_varing_r_el}. These errors are significantly smaller than the typical energy separations between adjacent band lines, attesting to the robustness of the model.

Once the electron-lattice cutoff has been established, we proceed to verify the convergence of the energy thresholds $\delta f$ and $\delta g$. For a fixed cutoff of $\delta r_{\rm el} = 4.0$~\AA, the residual errors seen in Fig.~\ref{fig:sto_bars_varing_r_el}(c) can be further reduced by setting $\delta f = 0.01$~eV/\AA\ and $\delta g = 0.01$~eV/\AA$^2$. These results demonstrate that systematic improvements in model accuracy can be achieved by tightening the cutoffs on the electron-lattice coupling terms.

Analysis of the error-bar plots reveals that the largest deviations occur near the $\Gamma$ point for both valence and conduction bands, and around the $R$ point for the conduction band. This behavior can be traced to the narrow energy range over which the $t_{\rm 2g}$ and $e_{\rm g}$ bands overlap in the RAG. As shown in Fig.~\ref{fig:sto-bands-windows}, the $t_{\rm 2g}$ bands at $R$ overlap energetically with the $e_{\rm g}$ bands at $\Gamma$. Although these manifolds exhibit weak entanglement in the high-symmetry RAG configuration, the octahedral rotations characteristic of the antiferrodistortive phase introduce mixing between the $\sigma$-like $e_{\rm g}$ and $\pi$-like $t_{\rm 2g}$ states. This enhanced entanglement leads to the increased errors observed near the $R$ point in Fig.~\ref{fig:sto_bars_varing_r_el}.
Since the discrepancies appear at relatively high energies—approximately 5 eV above the valence band maximum—their impact on the prediction of optical properties below the vacuum ultraviolet range, or on phenomena such as polaronic behavior, is expected to be minimal.
We have explored the possibility of mitigating this error by explicitly including the $e_{\rm g}$ bands in the model. This approach indeed improves the agreement between second-principles and DFT results in the energy range where the $e_{\rm g}$ and $t_{\rm 2g}$ manifolds overlap. However, due to the strong entanglement of the $e_{\rm g}$ states with higher-energy bands that are not incorporated into the model, this extension leads to large errors at elevated energies.
Therefore, we conclude that increasing the number of Wannier functions to include the $e_{\rm g}$ bands does not offer a viable path toward improving the overall accuracy of the model.

% ****************************

We now proceed to determine the electron-electron interaction parameters, $U_{\bm{ab},\bm{a^\prime b^\prime}}$ and $I_{\bm{ab},\bm{a^\prime b^\prime}}$, which quantify the variation of the Hamiltonian matrix elements $h_{\bm{ab}}$ under changes in the total electronic density (sum over both spin channels) and the spin polarization (difference between spin channels), respectively.
The standardized training set introduced in Sec.~\ref{sec:param} incorporates both electron and hole doping, as well as magnetized configurations. This allows us to probe how the occupancy of both titanium and oxygen orbitals affects the electronic structure and assess the material’s tendency toward spin polarization.

\begin{figure}[t]
    \centering
     \includegraphics[width=1.0\linewidth]{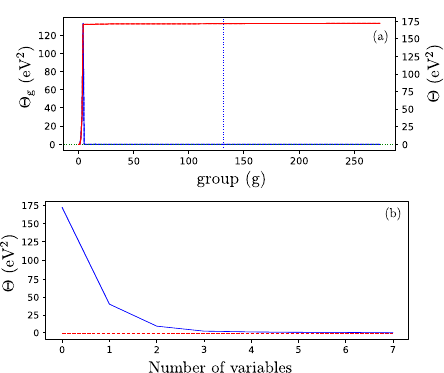} 
     \caption{\justifying (Color online) (a) Decomposition of the goalfunction for the electron-electron training set into contributions from each group of symmetry-equivalent $h_{\bm{ab}}$ matrix elements (blue line),  see Eq.~(\ref{eq:goalfunction_group}), along with the cumulative goalfunction $\Theta$ (red line). The dashed vertical line represents $ \delta r_{\rm h}$. (b) Evolution of $\Theta$ as electron-electron interaction parameters are progressively introduced into the second-principles model.}
    \label{fig:sto-ee}     
\end{figure}

To apply the fitting protocol described in Sec.~\ref{sec:train_ee}, we introduce a cutoff parameter, $\delta \Theta$, which defines the set of WF index pairs $\bm{a}\bm{b}$ to be included in the fitting of $U_{\bm{ab},\bm{a^\prime b^\prime}}$ and $I_{\bm{ab},\bm{a^\prime b^\prime}}$. Figure~\ref{fig:sto-ee}(a) displays the contribution of each symmetry-related group of interactions (g) to the electron-electron goalfunction calculated as,
\begin{align}
\Theta_{\rm g} = \sum_{\bm{ab}\in\text{g}} \sum_A \left[h_{\bm{ab}}(A)-h_{\bm{ab}}^\text{DFT}(A)\right]^2
\label{eq:goalfunction_group}
\end{align}
when the Hamiltonian does not include any electron-electron interaction parameter.
In this plot, the red line represents the total error contribution as a function of group index, ordered by increasing WF separation.
We observe a steep rise in the goalfunction for the initial groups, corresponding to interactions between closely located Wannier functions. This trend quickly saturates, indicating that the dominant electron-electron interactions are intra-atomic, while inter-atomic terms contribute significantly less. Notably, the strongest contributions stem from interactions involving Ti-centered 3$d$-like WFs, consistent with the localized nature of these orbitals.
Based on this analysis, we set the cutoff $\delta \Theta$ = 0.2 eV$^{2}$, as indicated by the green horizontal line in Fig.~\ref{fig:sto-ee}(a). This value captures the most significant variations in the Hamiltonian elements, specifically the diagonal terms $h_{\bm{aa}}$ involving Ti(3$d$), O(2$p_\sigma$), and O(2$p_\pi$) orbitals, as well as the intra-atomic off-diagonal interactions, $h_{\bm{ab}}$, between O(2$p_\sigma$) and O(2$p_\pi$).

Following the inclusion of $U$ and $I$ terms in the model, we observe a substantial reduction in the goalfunction quantifying the difference between DFT and second-principles results, as shown in Fig.~\ref{fig:sto-ee}(b). The most significant reduction—from approximately 170 eV$^2$ to 40 eV$^2$—is obtained by including the diagonal terms $U_{\bm{aa},\bm{aa}}$ and $I_{\bm{aa},\bm{aa}}$, where $\bm{a}$ corresponds to Ti ($t_{\rm 2g}$) WFs.
This reduction corresponds with the elimination of most part of the error $\Theta_g$ in the first group which is the largest peak in Fig.~\ref{fig:sto-ee}(a).
The second most impactful reduction, amounting to approximately 23 eV$^2$, results from incorporating similar diagonal terms for O(2$p_\pi$) orbitals. 
A third significant reduction is achieved by introducing intra-atomic off-diagonal terms between O(2$p_\pi$) and O(2$p_\sigma$) WFs. All remaining variables contribute less than 0.5 eV$^2$ to the goalfunction and can be safely neglected in the final model and are very difficult to observe in the scale shown in Fig.~\ref{fig:sto-ee}(a).

In the preceding paragraphs, we have demonstrated that a comprehensive electronic model for SrTiO$_{3}$, a prototypical transition-metal perovskite, can be constructed in a quasi-automated manner. The methodology presented here requires the user to select only a small set of threshold parameters, enabling the generation of accurate second-principles models with relatively modest computational resources.

The training set typically consists of a few hundred single-point calculations (on the order of 100–300) performed on a medium-sized supercell; in the present case, the supercell contains 40 atoms. When combined with a second-principles lattice model constructed following Refs.~\cite{wojdel_jpcm13,Escorihuela_prb17}, the result is a fully parameterized model capable of describing both structural and electronic properties.

Importantly, the current electronic model can be coupled with the recently developed real-time time-dependent second-principles DFT (SPDFT) method~\cite{tfr_prb25} to compute optical spectra. These predictions show excellent agreement with direct DFT calculations performed using the same exchange-correlation functional employed in the training.

A more challenging application is the description of polaron formation and dynamics in SrTiO$_{3}$. In this regard, we attempted to induce polaron formation by doping the system with electrons. However, these efforts did not result in charge localization. Benchmark DFT calculations on large supercells indicate that this failure is likely due to the lack of self-interaction correction in the LDA functional, which inhibits polaron formation.

To overcome this limitation, future work will focus on constructing second-principles models based on hybrid functionals. In particular, we plan to exploit the recent implementation of hybrid functionals in \textsc{siesta}~\cite{Garcia-20} to develop SrTiO$_{3}$ models trained using results obtained within the HSE06 approach~\cite{hse06}, which offers a more accurate treatment of electron localization and should enable a realistic description of polaronic effects.

\subsection{LiF}\label{sec:lif}

In this second example, we consider lithium fluoride (LiF), a prototypical strongly ionic compound whose optical spectrum is characterized by highly localized excitonic states~\cite{block_ssc78,williams_jpcs90,song_excitonbook}.
The presence of a pronounced Stokes shift~\cite{song_excitonbook} further indicates significant electron-phonon coupling in the excited state.
This unique combination of features makes LiF an ideal candidate for investigation within the second-principles framework, particularly using the recently developed time-dependent propagation formalism~\cite{tfr_prb25}. 
To enable such simulations, however, it is first necessary to construct an accurate second-principles model capable of describing the band structure, electron-lattice coupling, and electron-electron interactions across the valence and conduction manifolds.
In this case, the RAG corresponds to the face-centered cubic (FCC) structure with space group Fm$\bar{3}$m, characteristic of the rock-salt lattice. A conventional unit cell comprising eight atoms and a lattice constant of 4.026~\AA\ is employed for the simulations.

\begin{figure}[h]
    \centering
    \includegraphics[width=\linewidth]{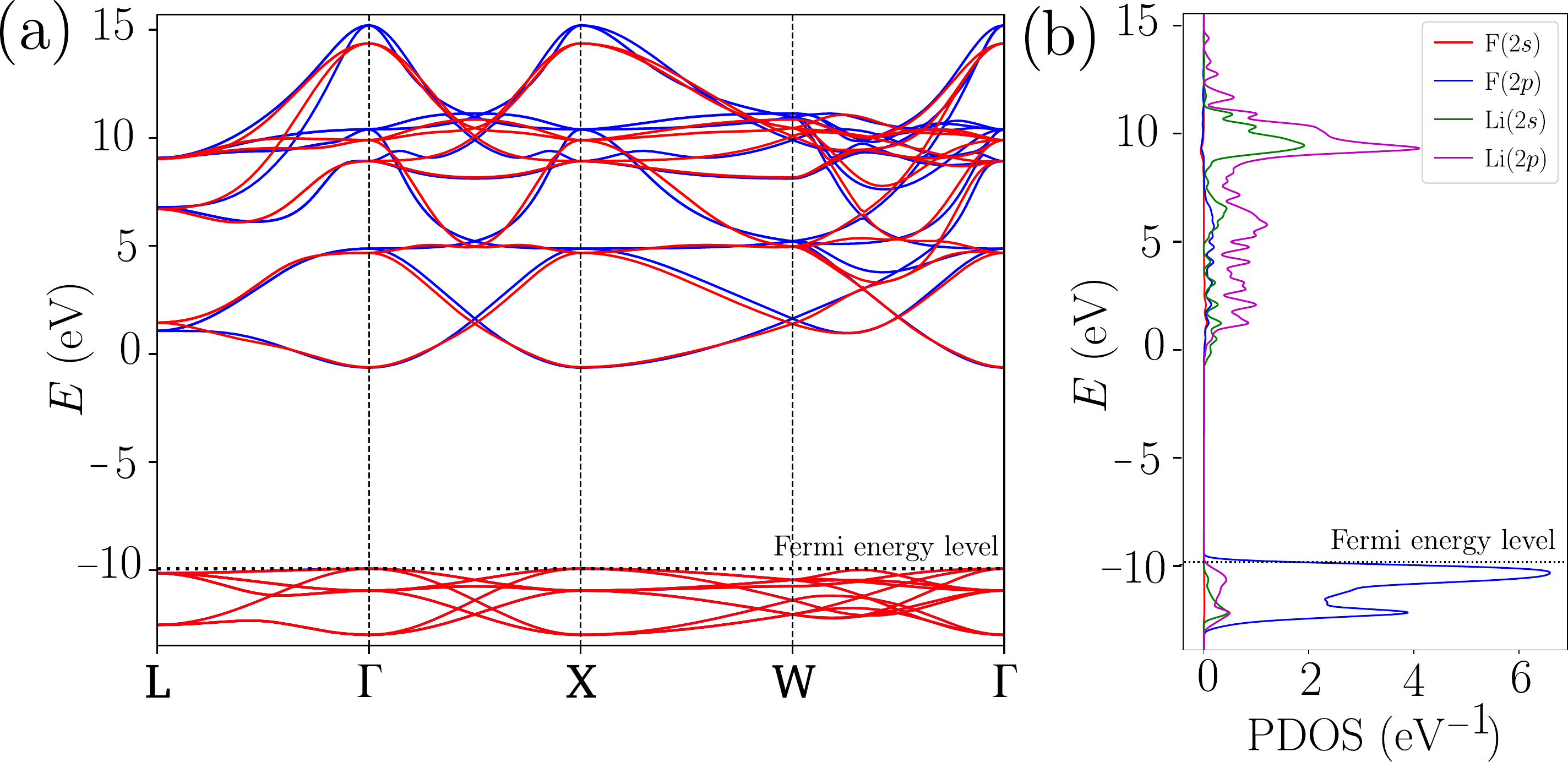} 
    \caption{\justifying (Color online) (a) Electronic band structure of LiF at the cubic RAG geometry, as obtained from DFT (blue) and second-principles simulations (red). (b) Corresponding projected density of states (PDOS), indicating that the valence bands are predominantly composed of F($2p$) orbitals, while the conduction bands exhibit mainly Li($2s$) and Li($2p$) character. In \textsc{siesta}, both the valence and conduction manifolds are disentangled from other states and used to construct the second-principles model.}
    \label{fig:lif-bands}
\end{figure}

Figure~\ref{fig:lif-bands} presents in blue color the electronic band structure of LiF, computed for the conventional unit cell used to train the second-principles model, employing the PBE density functional as implemented in \textsc{siesta}.
The valence bands, primarily of F($2p$) character, are relatively narrow, while the conduction bands are significantly more dispersive and exhibit dominant contributions from Li($2s$) and Li($2p$) orbitals (see note in Ref.
\footnote{%When comparing the band structure obtained with \textsc{siesta} to that produced by a plane-wave-based code, good agreement is observed for both the valence bands and the lower portion of the conduction manifold.
% 
%At higher energies, however, notable differences emerge: while the plane-wave calculation yields fully entangled conduction bands, the localized basis set used in \textsc{siesta} results in a spurious gap between band manifolds.
% 
%Although this separation is an artifact of the more limited basis set, the lower conduction bands remain reliable due to their close correspondence with the plane-wave results.
% 
%This separation is advantageous for model construction, as it simplifies the definition of a smooth and consistent Hamiltonian under perturbations such as structural deformations or doping.
% 
%In contrast, the use of disentanglement procedures can introduce discontinuities in the extracted Hamiltonian parameters, compromising their transferability.
% 
%Therefore, the naturally disentangled character of the conduction manifold in \textsc{siesta} represents a significant benefit for second-principles modeling, enabling a direct and robust evaluation of electron-lattice coupling parameters via finite-difference methods, as described in Sec.~\ref{sec:train_ellat}.
The low-lying conduction bands obtained in \textsc{siesta} are a good match for those of a more accurate plane-wave code. At higher energies, beyond the usual range of optical/UV experiments, the bands differ. In particular, the ones obtained with \textsc{siesta} become disentangled. This fortuitous event allows us to avoid sudden changes in the Wannier Hamiltonian due to changes in the disentaglement procedure.}).

To construct the Wannier functions (WFs) from the DFT data, we project the Bloch wavefunctions of the valence band onto F($2p$) atomic orbitals, and those of the conduction band onto Li($2s$) and Li($2p$) orbitals.
Following the procedure used for SrTiO$_{3}$, we employ the $C(r)$ function [Eq.~(\ref{eq:h_fun})] to determine a suitable cutoff distance for the Hamiltonian interactions, $\delta r_{\rm h}$, as illustrated in Fig.~\ref{fig:lif-ee}.
In this case, the spatial extent of the Hamiltonian matrix elements is somewhat larger than that found in SrTiO$_{3}$.
Nevertheless, adopting $\delta r_{\rm h} = 8.0$~\AA, as in the previous case, yields an accurate reproduction of the DFT band structure, represented in red in Fig.~\ref{fig:lif-bands}, with the exception of the uppermost conduction bands located approximately 20 eV above the valence band maximum.
These high-energy states lie well beyond the spectral range typically relevant for optical or ultraviolet studies, and also fall within the regime where the accuracy of \textsc{siesta}'s band structure is limited by the relatively small  basis set employed.

\begin{figure}[h]
    \centering
    \includegraphics[width=0.9\columnwidth]{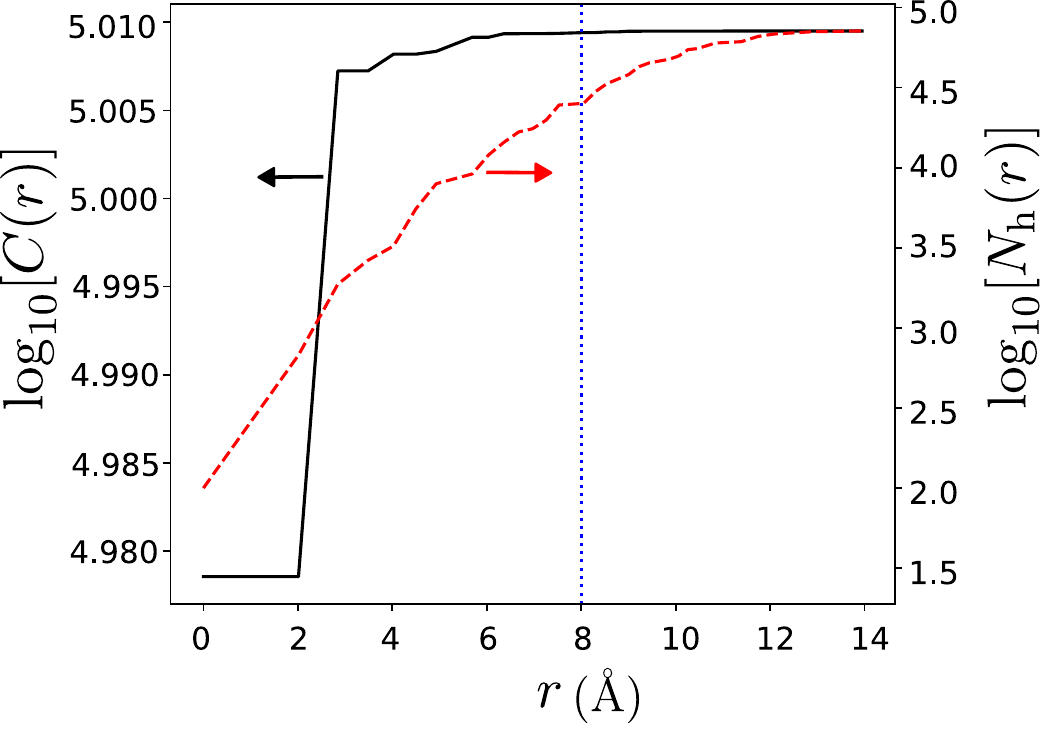} 
    \caption{\justifying Logarithmic plot of the $C(r)$ function (black, left axis) and the number of one-electron Hamiltonian matrix elements as a function of distance (red, right axis). The vertical dashed line indicates the chosen Hamiltonian cutoff distance, $\delta r_{\rm h} = 8$\AA, used to construct the second-principles model for LiF. This choice corresponds to the band structure shown with dashed red lines in Fig.~\ref{fig:lif-bands}(a).}
    \label{fig:lif-ee}
\end{figure}

\begin{figure*}
    \centering  
    \includegraphics[width=0.8\textwidth]{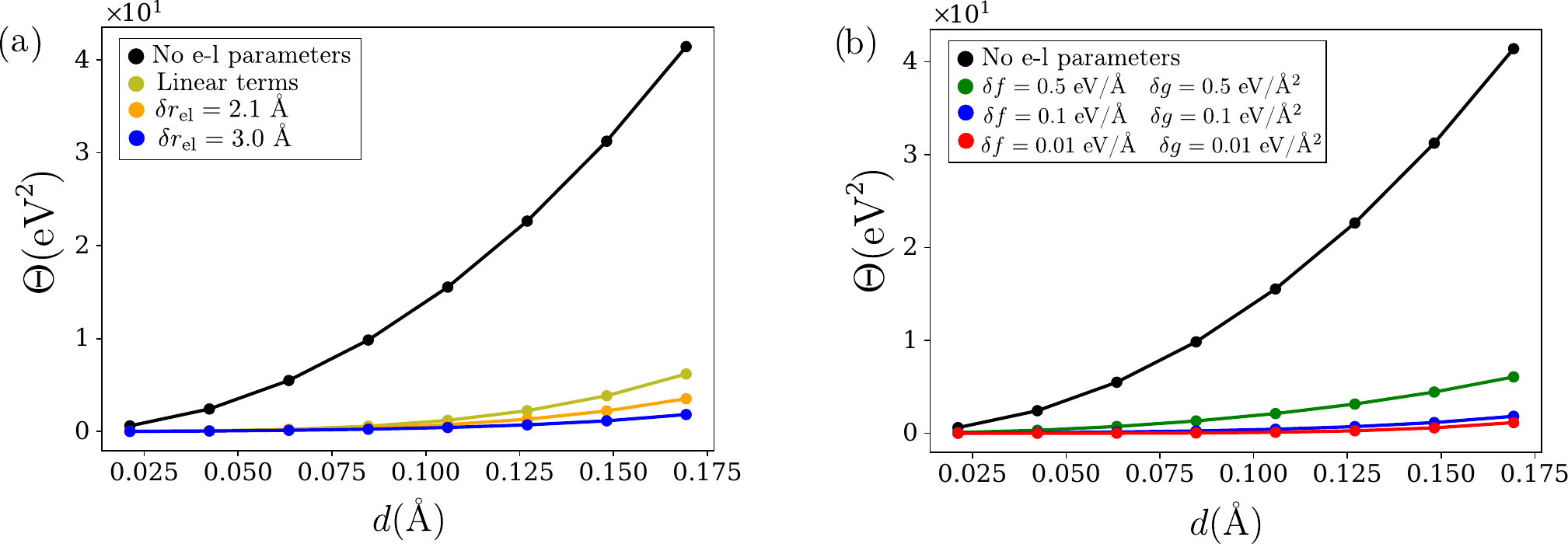}

     \caption{\justifying
    (Color online) Goalfunction $\Theta$ for LiF, measuring the average error per calculation in \textit{test set} comprising ten configurations with increasing atomic displacements $d$ from the reference atomic geometry (RAG). (a) The thresholds are fixed at $\delta f = 0.1$ eV/\AA\ and $\delta g = 0.1$ eV/\AA$^2$. The black line corresponds to a model excluding electron-lattice interactions; the olive green line includes only linear couplings. Models incorporating both linear and quadratic terms are shown in orange and blue, corresponding to cutoff distances of $\delta r_{\rm el} = 2.1$~\AA\ and $\delta r_{\rm el} = 3.0$~\AA, respectively. (b) The electron-lattice cutoff is fixed at $\delta r_{\rm el} = 3.0$~\AA, while different pruning thresholds are explored: $\delta f = 0.5$ eV/\AA, $\delta g = 0.5$ eV/\AA$^2$ (green), $\delta f = 0.1$ eV/\AA, $\delta g = 0.1$ eV/\AA$^2$ (blue), and $\delta f = 0.01$ eV/\AA, $\delta g = 0.01$ eV/\AA$^2$ (red).     
     }
     \label{fig:lif_ellat_cutoffs}
\end{figure*}

As in the case of SrTiO$_{3}$, we extend the applicability of the electronic model for LiF beyond the RAG by incorporating electron-lattice interactions. To this end, we analyze the sensitivity of the model to the electron-lattice cutoff distance, $\delta r_{\rm el}$, which determines the range of atomic pairs contributing to the electron-lattice coupling, as well as the thresholds $\delta f$ and $\delta g$ used to discard weak linear and quadratic interactions, respectively.
Figure~\ref{fig:lif_ellat_cutoffs}(a) shows the variation of the goalfunction as a function of the distortion amplitude $d$ (defined in Sec.~\ref{sec:STO}) for two different cutoff values, $\delta r_{\rm el} = 2.1$~\AA\ and 3.0~\AA. The former includes only first-neighbor Li–F interactions, while the latter also captures second-neighbor Li–Li and F–F interactions. Even for relatively large distortions ($d \approx 0.17$~\AA), characteristic of thermal motion at approximately 400 K, the total goalfunction remains low ($\Theta \approx 1$ eV$^2$), summed over 11964 terms. The model with $\delta r_{\rm el} = 3.0$~\AA\ yields an error roughly 60 \% smaller than that with $\delta r_{\rm el} = 2.1$~\AA. However, due to the overall small magnitude of the error, this improvement is of limited practical significance, indicating that the dominant contributions arise from short-range linear interactions.
To further reduce the model size, we assess the effect of pruning weak interactions based on the thresholds $\delta f$ and $\delta g$. In Fig.~\ref{fig:lif_ellat_cutoffs}(b), we evaluate the model accuracy for several threshold values. A noticeable increase in the goalfunction is observed when $\delta f = 0.5$~eV/\AA\ and $\delta g = 0.5$~eV/\AA$^2$, compared to the results obtained for $\delta f = 0.1$~eV/\AA, $\delta g = 0.1$~eV/\AA$^2$, and the more stringent $\delta f = 0.01$~eV/\AA, $\delta g = 0.01$~eV/\AA$^2$. These findings suggest, in line with what we observed in the SrTiO$_3$ case, that accurate and compact models can be constructed using $\delta f = 0.1$~eV/\AA\ and $\delta g = 0.1$~eV/\AA$^2$ as standard pruning thresholds.

\begin{figure*}[t]
    \centering
    \includegraphics[width=\linewidth]{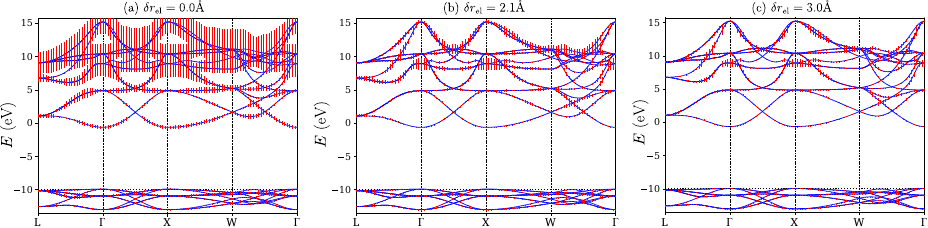}
    \caption{\justifying (Color online) Band structures illustrating the statistical error of the second-principles model across a test set of ten distorted geometries. The error bars (in red) represent the standard deviation of the energy levels relative to the reference DFT values. Panel (a) shows results obtained using a second-principles model without electron-lattice interactions. Panels (b) and (c) include linear and quadratic electron-lattice coupling terms with cutoff distances $\delta r_{\rm el} = 2.1$~\AA\ and $3.0$~\AA, respectively. All models were generated using thresholds $\delta f = 0.1$~eV/\AA\ and $\delta g = 0.1$~eV/\AA$^2$.}
\label{fig:lif_error_bars_varying_cfpair}
\end{figure*}

To assess the distribution of distortion-induced errors across the electronic band structure, we analyze the band-resolved deviations for atomic configurations with a maximum displacement of $d = 0.17$~\AA, as shown in Fig.~\ref{fig:lif_error_bars_varying_cfpair}. These results illustrate how the inclusion of electron-lattice interaction terms progressively improves the model accuracy as the cutoff parameter $\delta r_{\rm el}$ is increased. Figure~\ref{fig:lif_error_bars_varying_cfpair}(a) shows the error bars obtained in the absence of electron-lattice corrections, while Figs.~\ref{fig:lif_error_bars_varying_cfpair}(b) and \ref{fig:lif_error_bars_varying_cfpair}(c) correspond to models including electron-lattice interactions with $\delta r_{\rm el} = 2.1$~\AA\ and $\delta r_{\rm el} = 3.0$~\AA, respectively. The comparison indicates that the dominant contributions to the electron-lattice coupling arise from nearest-neighbor Li–F interactions, while the inclusion of second-neighbor Li–Li and F–F couplings yields a modest improvement in accuracy. Furthermore, we find that the valence bands and the lower part of the conduction manifold exhibit minimal errors across all cases. The largest deviations appear in the upper conduction bands; however, these states lie well above the vacuum ultraviolet range and are thus of limited relevance for most physical applications. Overall, the model demonstrates high reliability for the prediction of spectroscopic and dynamical properties in realistic conditions.

As in the case of SrTiO$_3$, the training set for LiF includes electron and hole doping at levels of 0.1, 0.2, and 0.3 electrons, both under spin-restricted (non-magnetic) and spin-polarized conditions. These configurations are used to fit the $U$ and $I$ parameters associated with the valence and conduction bands. Given the known importance of electron-hole interactions in accurately describing the optical properties of LiF, we further include spin-polarized calculations on the neutral system, constraining the total magnetization to 0.5, 1.0, 1.5, and 2.0 electrons. These configurations effectively promote electrons from the valence to the conduction band, enabling a more precise characterization of electron-hole coupling.
This comprehensive training strategy allows \textsc{modelmaker} to fit interaction terms of the form $U_{\bm{ab},\bm{a^\prime b^\prime}}$ and $I_{\bm{ab},\bm{a^\prime b^\prime}}$ that couple valence and conduction  Wannier functions. When plotting the total error function $\Theta$ for a model that excludes electron-electron interactions, we observe a pattern similar to that found in SrTiO$_3$: the dominant contributions to the error originate from a limited number of interaction groups, primarily associated with intra-atomic interactions [Fig.~\ref{fig:lif-ee}(a)]. However, in LiF, we additionally identify several groups corresponding to longer-range interactions that, while smaller, yield non-negligible contributions to $\Theta$.
As in the $\rm SrTiO_3$ case, we adopt a threshold of 0.2 eV$^2$ to filter relevant electron-electron terms. Beyond the expected intra-atomic contributions, we find that several F(2$p$)-F(2$p$) Hamiltonian matrix elements exhibit sensitivity to the electronic state. Fitting the $U$ and $I$ constants corresponding to these indices leads to a substantial reduction of the error function, as shown in Fig.~\ref{fig:lif-ee}(b).
The most significant improvements stem from three interaction variables: an intra-atomic interaction on Li [contributing 279 eV$^2$ and that essentially neutralizes the large $\Theta_g$ shown as a large peak in Fig.~\ref{fig:lif-ee}(a)], and two centered on F (together contributing 116 eV$^2$). Notably, we also identify a significant inter-atomic interaction involving $U_{\bm{aa},\bm{bb}}$ and $I_{\bm{aa},\bm{bb}}$ terms, where WF $\bm{a}$ has F(2$p$) character and WF $\bm{b}$ corresponds to a first-neighbor Li(2$s$) orbital. This interaction alone reduces $\Theta$ by approximately 4.5 eV$^2$, confirming the presence of non-negligible inter-site electron-hole coupling. All other terms contribute individually less than 0.8 eV$^2$ to the reduction of the error function.
These results confirm that the method successfully captures strong electron-hole interactions in LiF, in agreement with experimental observations of highly localized excitons in this material~\cite{song_excitonbook}.

\begin{figure}[h]
    \centering
    \includegraphics[width=\columnwidth]{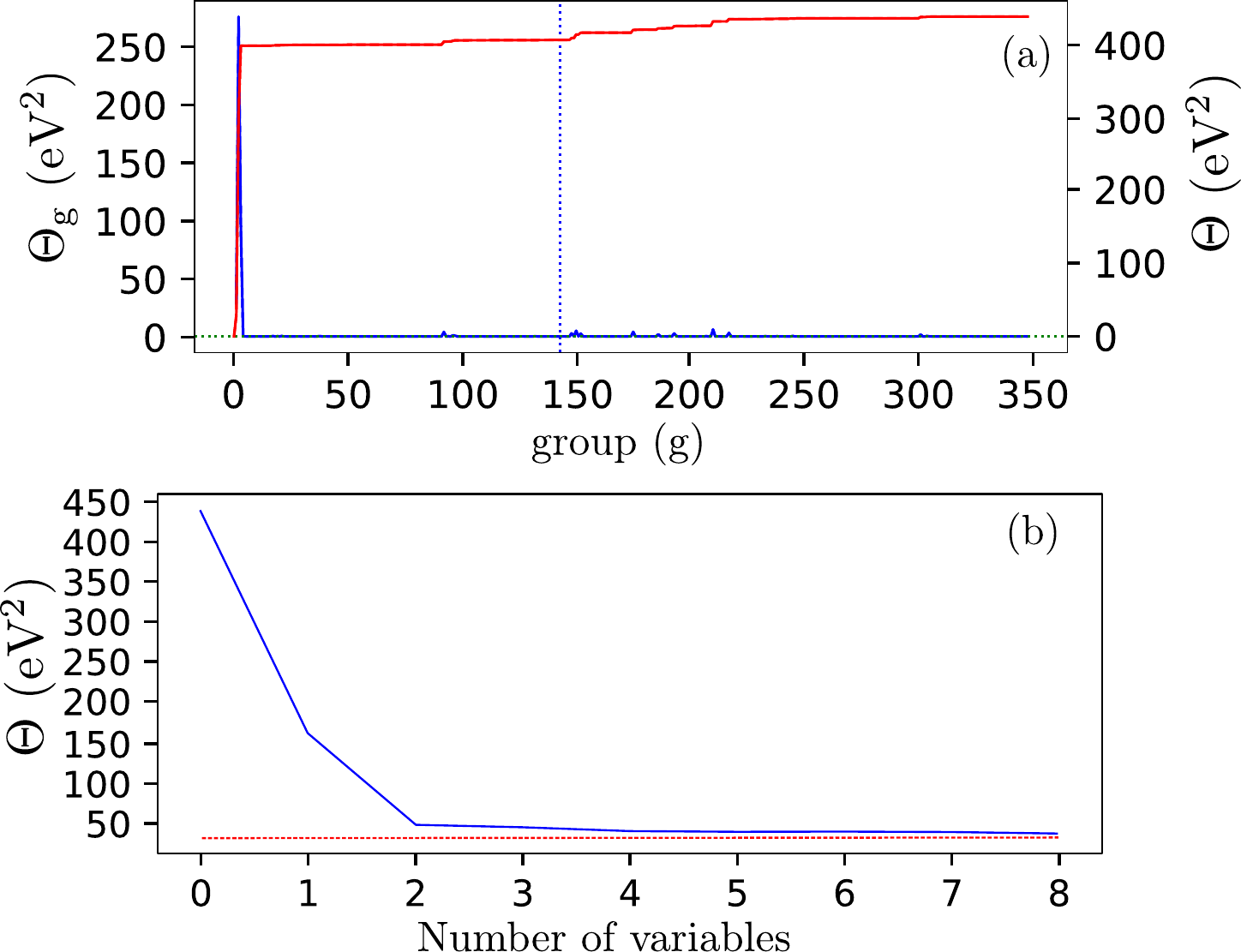} 
    \caption{\justifying (Color online) (a) Decomposition of the goalfunction for the electron-electron training set in LiF. The contributions from each group of symmetry---equivalent Hamiltonian matrix elements $h_{\bm{ab}}$--- are shown as a blue curve, $\Theta_{\rm g}$; while the cumulative goalfunction $\Theta$ is plotted in red. The dashed vertical line represents $\delta r_{\rm h}$. (b) Evolution of $\Theta$ as successive electron-electron interaction terms are incorporated into the second-principles model.}
    \label{fig:lif-ee}
\end{figure}

\section{Conclusions}
\label{sec:conclusions}

In this work, we have introduced a quasi-automated framework for constructing second-principles (SP) electron models, enabling efficient and physically grounded parameterizations of complex materials. Central to our approach is the use of carefully designed training sets—applied uniformly across materials—which allow the systematic control of model quality through a limited set of physically meaningful parameters. This procedure ensures both flexibility and comparability across systems.

The incorporation of symmetry throughout the model construction process significantly reduces the number of independent parameters and the number of required DFT calculations, yielding computational savings of several orders of magnitude. Validation against first-principles simulations reveals that the model accuracy depends primarily on the character of the electronic states and the magnitude of the applied perturbations. Nonetheless, the valence bands and the lower part of the conduction band—those most relevant for optical and transport phenomena—are consistently well reproduced, even under substantial structural distortions.

Compared to earlier formulations of SPDFT models~\cite{pgf_prb16}, this methodology represents a substantial advancement in generality, automation, and physical interpretability. More importantly, our goal extends beyond mere reproduction of DFT results: we aim to develop models that can reliably predict properties not included in the training data, including many-body interactions and finite-temperature effects.

To this end, we have developed a complementary real-time time-dependent SPDFT framework~\cite{tfr_prb25}, which allows the computation of dynamic observables such as currents induced by time-dependent electric fields. This formalism is well suited for the study of emergent quasiparticles—such as polarons and excitons—whose behavior stems from strong electron-lattice and electron-electron (or electron-hole) interactions. While the analysis of these effects is beyond the scope of the present article, the models generated here have already proven capable of describing optical phenomena in wide-gap materials such as LiF.  This will be discussed in an upcoming article.

The model construction algorithm is also amenable to future refinement. For instance, electron-lattice couplings can, in some cases, be directly computed from first-principles by analyzing the change in specific Hamiltonian matrix elements $h_{\bm{ab}}$ under targeted atomic displacements. However, analogous procedures to compute electron-electron interaction terms—such as the $U_{\bm{ab},\bm{a}'\bm{b}'}$—are not yet available, largely due to the difficulty in controlling charge density perturbations in a physically meaningful way. Progress in this direction could also inform improvements to DFT+$U$ methodologies and related approaches.

In the current context of model development, it is worth reflecting on the increasing popularity of machine-learning (ML) techniques. ML models have demonstrated impressive accuracy and efficiency, particularly in molecular dynamics and electronic structure predictions~\cite{schmidt_npj19,baum_jcim21}. However, such a complexity and ultimate absence of physical footing also limits their predictive power, particularly relative to physically motivated models. First, ML models are typically black boxes, lacking transparent traceability between inputs and outputs. Second, their parameters often lack direct physical interpretation, hindering microscopic insight. For example, our linear electron-lattice couplings $\vec{f}_{\bm{aa},\bm{\lambda}}$ directly quantify the force on a Wannier function $\chi_{\bm{a}}$ due to the motion of atom $\bm{\lambda}$, providing an intuitive connection between model and physics. Third, and most critically, ML models lack standardized metrics for transparency or model comparison. In contrast, our method uses clearly defined cutoff thresholds and hierarchical model construction, ensuring that expanded models systematically improve upon simpler ones—both in predictive power and physical interpretability.

Finally, we envision productive synergies between the type of structured, physics-based models developed here and ML techniques. The model files generated by \textsc{modelmaker} contain rich, well-structured datasets that could serve as ideal training inputs for ML models aimed at predicting a broader range of properties than currently feasible~\cite{schmidt_npj19}. Moreover, it is conceivable that new SP models could be generated through ML algorithms trained on existing parameter sets—especially in disordered systems such as interfaces, defects, or amorphous materials—where conventional fitting procedures are less tractable.  Ultimately, progress in this direction could results in physics-informed ML models for both lattice and electrons, potentially competitive with, and more transparent than, existing alternatives.

In summary, the framework presented here opens a path toward efficient, transparent, and physically insightful second-principles modeling. We anticipate that these tools will prove valuable in the prediction and interpretation of complex phenomena across a wide range of materials in the near future.

\begin{acknowledgements}
We acknowledge financial support from Grant No.~PID2022-139776NB-C63 funded by MCIN/AEI/10.13039/501100011033 and by ERDF/EU ``A way of making Europe'' by the European Union.
N.C.S. acknowledges financial support from ``Concepci\'on Arenal'' Grant No. BDNS:524538 of the University of Cantabria  funded by the Government of Cantabria.
T.F.R. acknowledges financial support from Ministerio de Ciencia, Innovaci\'on y Universidades (Grant PRE2019-089054). 
Also, J.Í-G. acknowledges the financial support from the Luxembourg National Research Fund through grant C21/MS/15799044/FERRODYNAMICS.
\end{acknowledgements}

\appendix

\bibliography{biblio}

\end{document}